\documentclass[sigconf, nonacm]{acmart}

\newcommand\vldbdoi{XX.XX/XXX.XX}

\newcommand\vldbavailabilityurl{https://github.com/maxsignalll/SAM}
\newcommand\vldbpagestyle{plain} 
\usepackage{algorithmic}
\usepackage{booktabs} 
\usepackage{makecell} 
\usepackage{tabularx}
\usepackage{graphicx}
\usepackage{float}
\usepackage{enumitem}
\usepackage{titlesec}
\usepackage[linesnumbered,ruled,vlined]{algorithm2e}
\titlespacing*{\section}{\parindent}{1.5ex}{1ex}
\titlespacing*{\subsection}{\parindent}{1ex}{0.5ex}
\titlespacing*{\subsubsection}{\parindent}{0.5ex}{0ex}
\titlespacing{\paragraph}{\parindent}{0.5ex plus 0.2ex minus 0.2ex}{0.5ex}
\begin{document}
\title{SAM: A Stability-Aware Cache Manager for Multi-Tenant Embedded Databases}

\author{Haoran Zhang}
\affiliation{%
  \institution{Harbin Institute of Technology}
  \city{Harbin}
  \state{Heilongjiang}
  \country{China}
}
\email{zhr@stu.hit.edu.cn}

\author{Decheng Zuo}
\affiliation{%
  \institution{Harbin Institute of Technology}
  \city{Harbin}
  \state{Heilongjiang}
  \country{China}
}
\email{zuodc@hit.edu.cn}

\author{Yu Yan}
\affiliation{%
  \institution{Harbin Institute of Technology}
  \city{Harbin}
  \state{Heilongjiang}
  \country{China}
}
\email{yuyan@hit.edu.cn}

\author{Zhiyu Liang}
\affiliation{%
  \institution{Harbin Institute of Technology}
  \city{Harbin}
  \state{Heilongjiang}
  \country{China}
}
\email{zyliang@hit.edu.cn}

\author{Hongzhi Wang}
\authornote{Corresponding author}
\affiliation{%
  \institution{Harbin Institute of Technology}
  \city{Harbin}
  \state{Heilongjiang}
  \country{China}
}
\email{wangzh@hit.edu.cn}

\begin{abstract}
The co-location of multiple database instances on resource constrained edge nodes creates significant cache contention, where traditional schemes are inefficient and unstable under dynamic workloads. To address this, we present \textbf{SAM(a Stability-Aware Manager)}, an autonomic cache manager that establishes \textit{decision stability} as a first-class design principle. It achieves this through its core control policy, \textbf{AURA(Autonomic Utility-balancing Resource Allocator)}, which resolves the classic exploitation-exploration dilemma by synthesizing two orthogonal factors: the \textbf{$\mathcal{H}$-factor}, representing proven historical efficiency (exploitation), and the \textbf{$\mathcal{V}$-factor}, for estimated marginal gain (exploration). Through this practical synthesis and adaptive control, SAM achieves sustained high performance with strategic stability and robustness in volatile conditions.

Extensive experiments against 14 diverse baselines demonstrate SAM's superiority. It achieves top-tier throughput while being uniquely resilient to complex workload shifts and adversarial workloads like cache pollution. Furthermore, its decision latency is highly scalable, remaining nearly constant as the system grows to 120 databases. Crucially, SAM achieves superior \textbf{decision stability}---maintaining consistent optimization directions despite noise, avoiding performance oscillations while ensuring predictable Quality of Service. These results prove that a principled, stability-aware design is essential for sustained high performance in real-world, large-scale systems.
\end{abstract}

\maketitle

\pagestyle{\vldbpagestyle}
\begingroup

\ifdefempty{\vldbavailabilityurl}{}{
\vspace{.3cm}
\begingroup\small\noindent\raggedright\textbf{PVLDB Artifact Availability:}\\
The source code, data, and/or other artifacts have been made available at \url{\vldbavailabilityurl}.
\endgroup
}
\section{Introduction}\label{introduction}
Embedded databases, such as SQLite~\cite{sqlite}, are integral to the modern computing ecosystem. Unlike traditional client-server (C/S) databases~\cite{postgresql, mysql}, they are designed as libraries to be directly integrated into an application. This architecture excels in scenarios requiring local, high-performance, and zero-management data storage, including mobile applications~\cite{android-sqlite}, IoT devices~\cite{sqlite-iot}, and metadata management in large-scale distributed systems~\cite{rocksdb}. Statistically, embedded databases are the most widely deployed database engines~\cite{sqlitenmost}, rendering it nearly ubiquitous.

However, the increasing complexity of systems has given rise to a critical challenge: the concurrent operation of multiple, independent, database-reliant services on a single host, a practice known as \textbf{multi-tenant co-location}, where each service and its database acts as a distinct tenant. An edge computing node, for example, may simultaneously run services for network monitoring, data acquisition, and local analytics, each employing its own SQLite database. Similarly, in cloud-native environments, a physical machine may host dozens of containerized tenants that use SQLite for state management, a pattern increasingly advocated by modern cloud platforms for deploying scalable, distributed services \cite{cloudflare-d1, flyio-sqlite-tenant}.

In such multi-tenant co-location environments, a significant performance 
bottleneck emerges from memory contention—specifically, for the cache of 
each database instance. The conventional approach of statically assigning 
a fixed cache size to each instance is profoundly inefficient under 
dynamic workloads~\cite{mehta_dewitt_vldb93}. An idle instance's cache lies fallow, wasting memory, while an instance experiencing a sudden load spike is starved for resources, leading to unpredictable performance swings that prevent sustained high throughput.

This inefficiency is particularly critical in modern \textbf{Cloud-Edge-Device} 
architectures. Here, cache misses trigger expensive network requests to 
remote data centers rather than local disk accesses, resulting in 
latencies that are orders of magnitude higher. Consequently, poor cache management creates volatile performance patterns, making sustained high performance impossible and user experience unpredictable.

Addressing this bottleneck is complicated by the content-dependent and 
non-linear nature of cache performance~\cite{zhang2020parrot,berger2018robinhood}, 
which creates a complex optimization problem with inherently conflicting objectives:
\begin{itemize}[nosep]
    \item \textbf{(C1) Workload Dynamism:} The system must be agile enough to swiftly reallocate cache to new, emerging hotspots as workloads shift~\cite{chen2020hotring}.
    \item \textbf{(C2) Cache Pollution:} It must be discerning enough to prevent ``cache polluters''---instances running cache-unfriendly workloads like full table scans that consume vast amounts of memory for little gain---from evicting more valuable data~\cite{mutlu2012eafcache, mittal2017survey}.
    \item \textbf{(C3) SLA Guarantees:} It must ensure that low-traffic but high-priority instances are not starved of resources by busier, less-critical tenants~\cite{berger2018robinhood}.
    \item \textbf{(C4) Decision Stability:} Finally, it must maintain consistent optimization decisions despite transient noise (e.g., network jitter) while remaining responsive to genuine workload shifts, as performance oscillations severely degrade sustained throughput~\cite{yang2022cachesack}.
\end{itemize}

These requirements often pull in opposite directions—agility conflicts with stability, protection conflicts with efficiency. Addressing these conflicting requirements in unison necessitates a strategy that considers multiple orthogonal factors. As a structural foundation, we adapt the mature \textbf{``reservation plus elasticity''} model from cloud computing, which we instantiate in our system as a \textbf{two-pool architecture}~\cite{liakopoulos2018no, al-dhuraibi2017elasticity}. This design provides each tenant with a guaranteed resource baseline via a dedicated \textbf{Fixed Pool} (the reservation) while enabling dynamic adjustments from a shared \textbf{Elastic Pool} (the elasticity). This architecturally provides a direct solution for \textbf{(C3) SLA Guarantees} and creates the foundation for addressing the remaining challenges through principled control of the elastic pool.

However, the model itself is incomplete, as a naive policy for the elastic pool fails to resolve the remaining dynamic challenges. A purely \textit{reactive} policy may struggle with \textbf{(C1) Workload Dynamism}, while a purely \textit{opportunistic} one is vulnerable to \textbf{(C2) Cache Pollution} and can induce severe performance \textbf{oscillations (C4)}. Therefore, the core problem this paper addresses is the design of an adaptive control policy for this elastic pool that achieves sustained high performance in volatile environments. Our key insight is that such performance requires decision stability as its foundation, which can only be achieved by explicitly navigating the fundamental trade-off between rewarding proven, historical efficiency (\textbf{exploitation}) and seeking future performance improvements (\textbf{exploration}).

To effectively balance this exploitation-exploration trade-off, we propose \textbf{SAM (a Stability-Aware Manager)}, a closed-loop control system whose design directly addresses these challenges (Fig.~\ref{fig:arch}). First, to ensure (C3) SLA Guarantees, SAM's two-pool allocation model provides deterministic baseline resources via a \textit{fixed pool}. Second, to tackle the dynamic challenges, SAM relies on its principled core control policy, \textbf{AURA(Autonomic Utility-balancing Resource Allocator)}, to manage the \textit{elastic pool}. Through a practical synthesis of two orthogonal factors, SAM prevents (C2) Cache Pollution by rewarding historically stable efficiency (\textbf{$\mathcal{H}$-factor}), while adapting to (C1) Workload Dynamism by estimating marginal gain (\textbf{$\mathcal{V}$}). Finally, to guarantee (C4) Decision Stability, SAM utilizes AURA's stability-enforcing actuation layer. By unifying these design principles, SAM achieves sustained high performance, combining strategic stability with robustness against external disturbances.

The main contributions of this paper are threefold as follows.
\begin{enumerate}[nosep]
    \item We propose \textbf{SAM}, a cache management system built on a novel, stability-first design principle. Its \textbf{two-pool architecture} is the first to systematically decouple baseline SLA guarantees from dynamic performance optimization, architecturally separating safety from efficiency.

    \item We introduce a \textbf{principled engineering approach} to stability-aware control, embodied in our policy, \textbf{AURA}. It navigates the complex exploitation-exploration trade-off by practically synthesizing a tenant's historical efficiency ($\mathcal{H}$-factor) with its forward-looking marginal gain ($\mathcal{V}$-factor).

    \item We provide extensive experimental validation of SAM's superiority and practical viability against 14 diverse baselines. Our results on standard, adversarial, and TPC-C workloads demonstrate near-optimal throughput, unique resilience, and excellent scalability.

\end{enumerate}

The remainder of this paper is organized as follows. Section~\ref{section2} presents the system model and problem formulation. Section~\ref{sec:architecture} details the SAM system's architecture and design philosophy. Section~\ref{sec:algorithm} presents the AURA control policy in detail. Section~\ref{sec:analysis} provides a theoretical analysis of the algorithm's stability properties and its computational complexity. Section~\ref{sec:evaluation} describes our experimental setup and evaluation. Section~\ref{sec:related_work} reviews related work, and Section~\ref{sec:conclusion} concludes.

\section{Problem Definition}\label{section2}

In this section, we formalize the problem of cache management for multi-tenant embedded databases. We define the system architecture, the performance metrics of interest, and the multi-objective optimization problem that our framework, SAM, is designed to solve.

\subsection{System Model}\label{systemmodel}
Our system model comprises a set of $N$ independent database instances, $D=\{d_{1},...,d_{N}\}$, co-located on a single host. These instances compete for a shared, finite total cache budget, $C_{total}$, which is measured in pages. The core problem is to dynamically determine the cache allocation for each instance, $a_i(t)$, at each time interval $t$, such that $\sum_{i=1}^{N}a_{i}(t) = C_{total}$. The performance of each instance is a function of its allocation, and we monitor key metrics such as its workload intensity (Operations per Second, \textit{ops}$_i$) and its cache effectiveness (Hit Rate, $HR_i$).

\subsection{Problem Formulation}
The cache allocation problem in our multi-tenant embedded databases environment is fundamentally a multi-objective optimization problem: an ideal allocation strategy must maximize system throughput while ensuring tail latency remains below specified SLA thresholds. This multi-objective goal can be formalized as maximizing a utility function $U(\mathbf{TPS}, \mathbf{L_{p99}})$, but directly optimizing such a complex function in a noisy, online setting is intractable.

We therefore posit that \textbf{Aggregate Effective Throughput} is a powerful and stable \textbf{proxy metric} for this overarching goal. In the Cloud-Edge-Device architectures we target, the latency gap between a cache miss ($L_{miss}$) and a hit ($L_{hit}$) is typically several orders of magnitude~\cite{zhang2024apecache, nsf2020latency}. As the average query latency can be modeled as $L_{avg, i} \approx HR_i \cdot L_{hit} + (1 - HR_i) \cdot L_{miss}$, maximizing the throughput-weighted sum of hit rates is the most direct way to minimize system-wide latency.

This leads to the following optimization problem: At each time step $t$, our goal is to find an allocation plan $a(t) = \{a_1(t), ..., a_N(t)\}$ that solves the following constrained optimization problem:

\vspace{-3mm}
\begin{equation}
\begin{aligned}\label{def:problem}
& \max_{a(t)}
& & \sum_{i=1}^{K} \left( \textit{ops}_i(t) \times HR_i\left(a_i(t)\right) \right) \\
& \text{subject to}
& & \sum_{i=1}^{K} a_i(t) = C_{\text{total}}, \\
& & & a_i(t) \ge \ell_i, \quad \forall i \in \{1, ..., K\}.
\end{aligned}
\end{equation}

where $C_{\text{total}}$ is the total available cache and $\ell_i$ is a configurable, per-tenant minimum allocation requirement to satisfy potential SLA guarantees.

\noindent\textbf{Problem Hardness and Approach.} The problem formulated in equation~\ref{def:problem} is computationally hard. Even in its offline form, it is a variant of the \textbf{Multiple-Choice Knapsack Problem (MCKP)}, which is NP-hard~\cite{kleinberg2006algorithm}. The complexity is exacerbated in our setting, which is both \textbf{online} (future workloads are unknown) and \textbf{black-box} (performance functions are unknown a priori). Therefore, in alignment with standard practice for complex systems research, we design \textbf{SAM}, an autonomic control framework that employs a control-theoretic learning-based policy to find near-optimal allocations in this dynamic environment~\cite{Mao2016DeepRM}.

\subsection{Conflicting Design Goals}
The design of an effective allocation strategy is complicated by several fundamental trade-offs, which directly correspond to the challenges outlined in Section~\ref{introduction}.

\textbf{Exploitation vs. Exploration.} This trade-off is central to addressing both \textbf{Challenge 1 (Workload Dynamism)} and \textbf{Challenge 2 (Cache Pollution)}. An algorithm that solely \textit{exploits} the current state may achieve high current throughput but risks missing emerging hotspots. Conversely, aggressive \textit{exploration} by chasing naive metrics like miss count can be deceived by cache-unfriendly workloads. As detailed in Section~\ref{sec:algorithm}, our system, SAM, resolves this through its core control policy, AURA, which practically synthesizes an exploitation signal (the \textbf{$\mathcal{H}$-factor}) with an exploration signal (the \textbf{$\mathcal{V}$-factor}).

\textbf{Global Throughput vs. Individual SLAs.} This conflict lies at the heart of \textbf{Challenge 3 (SLA Guarantees)}. A myopic strategy focused solely on maximizing aggregate throughput would starve low-traffic but high-priority instances. SAM's \textbf{two-pool architecture} (Section~\ref{sec:architecture}) directly resolves this conflict by separating resources for baseline SLA guarantees from those for dynamic optimization.

\textbf{Responsiveness vs. Decision Stability.} This is the core trade-off for \textbf{Challenge 4 (Decision Stability)}, especially in noisy edge environments. Overly reactive strategies are vulnerable to transient noise, causing performance oscillations that degrade sustained throughput. Conversely, excessive damping may miss genuine workload shifts. Our system, SAM, is designed to maintain decision consistency while preserving necessary adaptiveness, utilizing the stability-aware mechanisms within its AURA policy, such as its \textbf{Saturation Confidence} model and \textbf{momentum-based allocator} (Section~\ref{sec:algorithm}).

\subsection{System Model and Operating Principles}
Our framework is designed based on the following principles, which we argue enhance its generality and practical relevance.

\textit{Orthogonality to Replacement Policy.}
Our framework focuses on dynamically \textit{sizing} the cache allocated to each instance, decoupling our resource allocation logic from the internal page \textit{replacement} policy (e.g., LRU, LFU) of the database engine. This is a standard \textbf{separation of concerns} that mirrors real-world database administration, where cache sizing is a configuration concern independent of the engine's internal algorithms~\cite{chou1986evaluation}. This principle makes our framework general and readily deployable on top of any embedded database.

\textit{Black-Box Metric Observability.}
We assume that macroscopic performance metrics (e.g., ops, cache hits, misses) for each database instance are observable via an external interface. This is a practical assumption grounded in the design of modern database systems, which universally provide such observability for monitoring and administration through interfaces like `PRAGMA' commands in SQLite~\cite{sqlite} or status views in systems like MySQL and PostgreSQL~\cite{mysql, postgresql}. This enables our non-intrusive, external orchestrator design.

\textit{Network-Dominant Latency at the Edge.}
Our work specifically targets the performance challenges in modern \textbf{Cloud-Edge-Device} architectures~\cite{cloudflare-d1, flyio-sqlite-tenant}. In such distributed systems, the primary source of user-perceived tail latency is not local disk I/O---which often occurs on fast NVMe SSDs at the edge---but the high and variable latency of network access required to fetch data from a regional data center during a cache miss. With network round-trips being typically two to three orders of magnitude slower than local storage access~\cite{nsf2020latency, zhang2024apecache}, maximizing cache hits is the core optimization strategy. Other latency sources, such as lock contention, are considered orthogonal to the cache sizing problem addressed here.

\section{System Overview}\label{sec:architecture}

To address the constrained optimization problem defined in Section~\ref{section2}, we propose \textbf{SAM}, an autonomic cache management framework. The core of our \textbf{stability-aware} design philosophy is that principled, internal \textbf{stability} is the foundation for achieving both high performance and external \textbf{robustness} in volatile, multi-tenant environments. SAM puts this philosophy into practice through its core control policy, \textbf{AURA}, which moves beyond simple reactive heuristics through a principled synthesis of historical efficiency (\textbf{$\mathcal{H}$-factor}) and future potential (\textbf{$\mathcal{V}$-factor}).

This section details the concrete architecture of this framework. We begin by outlining its overall structure and the core two-pool resource model, then delve into the design principles behind its autonomic control policy.

\subsection{Overall Architecture}
The SAM system is designed to operate as an external, non-intrusive coordinator for a set of co-located, black-box database instances. As illustrated in Figure~\ref{fig:arch}, the architecture is composed of two primary conceptual layers. The \textbf{Resource Layer} partitions the \texttt{Total Host Cache} ($C_{total}$) into a \textbf{Fixed Pool} for static baseline guarantees, and an \textbf{Elastic Pool} for dynamic optimization.

The \textbf{System Layer} contains the core control logic. A centralized \textbf{SAM Autonomic Coordinator}, which implements the \textbf{AURA} control policy, manages the database instances residing in the \textbf{Managed Environment}. This architecture implements a classic \textbf{feedback control loop} operating in a \textit{Sense-Decide-Act} cycle:

\textbf{Sense:} The Coordinator gathers fine-grained \texttt{Performance Metrics} directly from each database instance, while also perceiving global \texttt{System Info} from the resource layer to maintain a comprehensive view of the system state.
\textbf{Decide:} In the decision phase, the Coordinator executes its AURA control policy, synthesizing the perceived state to calculate an optimal allocation plan. This plan, detailed in Section~\ref{sec:algorithm}, strategically balances historical performance with future potential.
\textbf{Act:} The Coordinator dispatches direct \texttt{Cache Resizing Commands} to each target instance, which then adjusts its memory footprint by requesting resources from or releasing them to the Elastic Pool.

This interaction is mediated by a set of simple, abstract interfaces. The Coordinator assumes only that a target instance provides a way to get metrics (e.g., `\texttt{get\_metrics()}') and a way to configure its cache size (e.g., `\texttt{set\_cache\_size()}'). This separation of concerns enhances the framework's generality, allowing it to be adapted to any embedded database that exposes similar observability and configurability.

\begin{figure}[t]
\centering
\includegraphics[width=\linewidth]{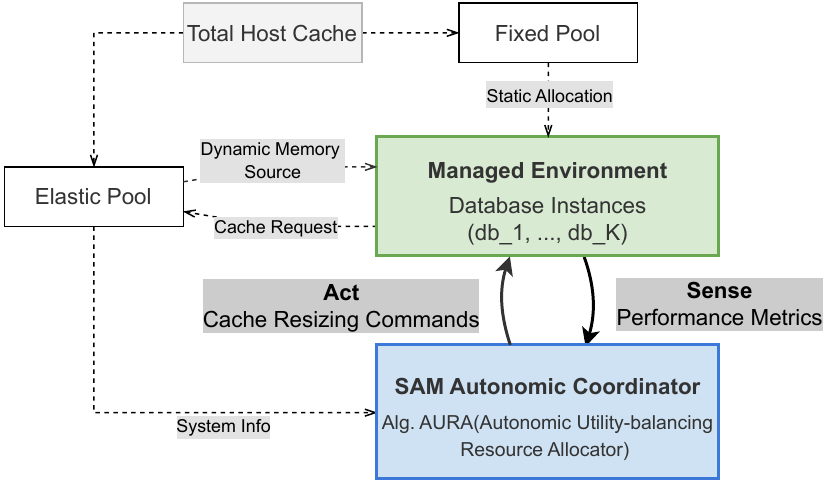}
\caption{Architecture of the SAM system, featuring a two-pool (Fixed/Elastic) resource model where a central Coordinator manages the elastic pool via a Sense-Decide-Act feedback loop.}
\vspace{-4mm}
\label{fig:arch}
\end{figure}

\subsection{The Two-Pool Cache Architecture}
A key architectural principle of our framework is the \textbf{two-pool cache architecture}, designed to resolve the inherent conflict between guaranteeing individual SLAs and optimizing for global performance. It operates by partitioning the total cache budget, $C_{total}$, into two distinct, non-overlapping pools. This decomposition is justified by the principle of \textit{separation of concerns}: the problem of satisfying hard performance constraints is architecturally separated from the problem of opportunistic, best-effort optimization.

\textbf{Fixed Pool ($C_{fixed}$).} 
This pool embodies our architectural principle that deterministic performance baselines for critical instances should be guaranteed through static allocation, not pursued as a goal within the dynamic optimization loop. This guarantee is realized through a two-step, policy-driven allocation process. First, an administrator defines the total size of the fixed pool, $C_{fixed}$, based on overall system service-level objectives. Second, this pool is apportioned among instances ($d_j \in D_p$) proportionally to their static, pre-defined, manually configured \texttt{base\_priority} values. The manually configured Fixed Pool is intentional, separating high-level policy set by administrators from the low-level, dynamic optimization handled by AURA. This static allocation, $a_j^{\text{fixed}}$, insulates critical instances from the resource volatility of the wider system. Critically, this same separation of concerns allows the manager for the remaining \textbf{Elastic Pool} to focus single-mindedly on maximizing overall system utility, without being constrained by the safety requirements of individual tenants. 

\textbf{Elastic Pool ($C_{elastic}$).} This pool contains all remaining cache resources ($C_{elastic} = C_{total} - C_{fixed}$) and is dedicated to maximizing the aggregate system utility, as defined by our objective function. The resources in this pool are entirely fluid and are dynamically reallocated in each tuning interval by the \textbf{SAM system}, guided by its core AURA control policy (which will be introduced in the next section). Finally, total allocation for any instance $d_i$ is the sum of its parts: $a_i^{\text{total}}(t) = a_i^{\text{fixed}} + a_i^{\text{elastic}}(t)$.

\subsection{The SAM Coordinator and the AURA Control Policy}
The autonomic control policy of the SAM framework resides in the control policy used by the Coordinator to manage the Elastic Pool. A simple policy, such as a greedy heuristic, is insufficient to navigate the conflicting challenges of dynamism, pollution, and stability defined in Section~\ref{section2}. Therefore, a principled, multi-factor control policy is required.

The design of our control policy is guided by two foundational principles. First, it is based on a \textbf{dual-factor utility paradigm} that resolves the exploitation-exploration trade-off by evaluating each instance along two orthogonal dimensions: its proven historical contribution (exploitation) and its marginal potential for future improvement (exploration). Second, it treats \textbf{stability as a first-class citizen}. Here, we distinguish between \textit{decision stability}—maintaining consistent optimization directions despite noisy observations—and mere \textit{response damping} that would slow adaptation. Our policy employs control-theoretic mechanisms to achieve the former, dampening internal oscillations while ensuring robust adaptations against external disturbances.

The concrete implementation of this control policy is our \textbf{AURA} engine, whose detailed specification is the subject of the next section. By effectively integrating these principles, this policy equips SAM with the ability to make stable, robust, and high-performance decisions, overcoming the limitations of simpler baselines.

\begin{figure*}[!htbp]
\centering
\includegraphics[width=\linewidth]{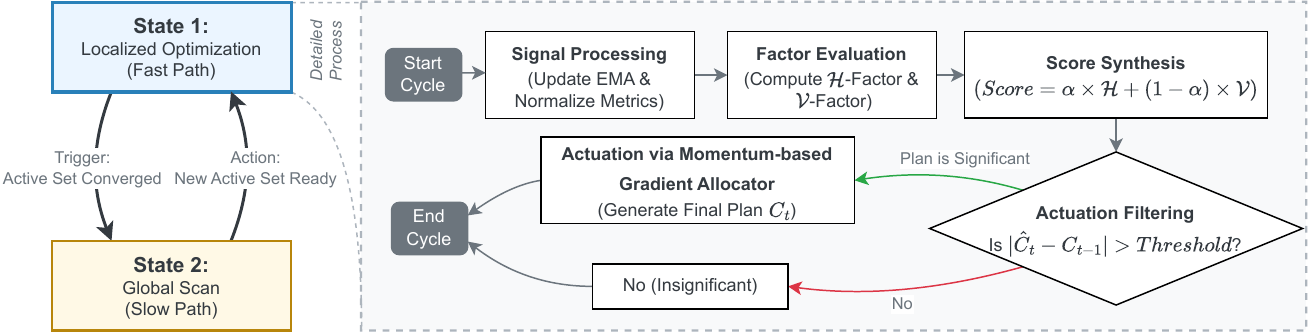}
\caption{The two-level architecture of the AURA control policy, showing: (a) the macro-level state machine that transitions between a fast-path \textit{Localized Optimization} and an infrequent \textit{Global Scan} to ensure scalability, and (b) the micro-level decision pipeline executed within the fast path.}
\label{fig:algo-arch}
\vspace{-3mm}
\end{figure*}

\section{The AURA Control Policy}\label{sec:algorithm}

The design principles articulated in the previous section are embodied in our core control policy, \textbf{AURA}, which serves as the control core of the SAM framework. This policy is realized through a two-level, adaptive state management architecture that maintains computational efficiency suitable for large-scale deployments. Figure~\ref{fig:algo-arch} provides a visual schematic of this architecture.

\subsection{Architectural Principles of the AURA}
As established in Section~\ref{section2}, the online, black-box nature of our problem makes finding a provably optimal solution intractable. Therefore, our approach is to design an autonomic control policy, AURA, to govern the SAM framework. This policy navigates the fundamental trade-off between exploitation and exploration to achieve the stability and robustness necessary for sustained high performance. It materializes this principle by synthesizing two orthogonal signals—a database's proven, historical efficiency and its potential for future performance gains—allowing SAM to make sound decisions in complex, dynamic environments.

At a macro-level, as shown in \textbf{Figure~\ref{fig:algo-arch}}, AURA employs an \textbf{Adaptive Active Set (AAS)} framework, which operates as a two-state machine. AURA spends the majority of its cycles in the computationally efficient \textbf{State 1 (Localized Optimization)}, focusing only on a small subset of tenants ($k \ll K$) to make rapid, high-quality decisions. It transitions to the more comprehensive \textbf{State 2 (Global Scan)} only when a strategic opportunity for improvement is detected (i.e., the active set converges). This design decouples decision latency from the system scale, providing excellent scalability while maintaining decision quality.

At a micro-level, \textbf{Figure~\ref{fig:algo-arch}} details the multi-stage decision pipeline executed within the fast-path (State 1). This pipeline is centered on the dual-factor scoring model through \textbf{Signal Synthesis}. It evaluates each instance using a dual-factor scoring system that balances the historical efficiency ($\mathcal{H}$-factor) with the future potential ($\mathcal{V}$-factor). The recommended plan generated by this process is then passed through a final \textbf{Actuation Filtering} stage. If deemed significant, the plan is translated into action by a \textbf{stability-enforcing actuator}---our stateful, \textbf{momentum-based gradient allocator}---which acts as a ``shock absorber'' to ensure all reallocations are smooth and deliberate.

In the following subsections, we will dissect each of these components in detail, from the core factor calculations to the state transition logic.

\subsection{Algorithm Specification}
We now provide the complete specification of the \textbf{AURA} control policy, translating the high-level architecture from the previous section into a concrete implementation. We begin by laying out the necessary data structures and state variables, followed by a detailed walkthrough of the main decision workflow and its innovative components as visually summarized in Figure~\ref{fig:algo-arch}.
\subsubsection{Data Structures and State.}
The AURA policy maintains state at two levels. The \textbf{Global Coordinator State ($S_g$)} includes: \textit{(i) Active Set State}, which contains the \texttt{current\_active\_set} of tenants being optimized and cached candidate pools (\texttt{top\_candidates}, \texttt{bottom\_candidates}); \textit{(ii) Convergence State}, such as a \texttt{score\_impro-\\vement\_window} and an \texttt{inactivity\_timer} for stagnation detection; and \textit{(iii) Global Policy State}, primarily the smoothed adaptive weight, \texttt{alpha\_prev}. At the per-instance level, the \textbf{Per-Instance State ($d_i$)} tracks: \textit{(a) Metric State}, including smoothed inputs like \texttt{ema\_ops\_slow}; \textit{(b) $\mathcal{V}$-Factor State}, such as \texttt{last\_allocation\_pages} to compute gradients; and \textit{(c) Stability Control State}, which features the crucial \texttt{saturation\_confidence} score.

\subsubsection{The Main Decision Workflow.}\label{sec:workflow}
AURA's core logic is architected as an adaptive, two-state workflow that dynamically focuses computation where it is most needed, ensuring high-quality decisions with excellent computational efficiency. The high-level process is visually summarized in Figure~\ref{fig:algo-arch} and specified in Algorithm~\ref{alg:sam_main}.

AURA's autonomic control policy employs an adaptive two-state architecture for computational efficiency. At the start of each cycle, a key decision is made (Algorithm~\ref{alg:sam_main}, Line~\ref{line:convergence_check}): should the system enter the infrequent, comprehensive \textbf{Global Scan} state, or continue in the lightweight, default \textbf{Localized Optimization} state? This decision is governed by a convergence metric that determines when the optimization potential of the current active set has been exhausted. This is achieved by monitoring a sliding window of performance score improvements; if the improvement rate stagnates, convergence is triggered.

If a global scan is triggered (e.g., on a cold start or after convergence), the system performs a highly efficient, linear-time pass using a \textbf{two-way heap filter} to gather fresh candidates (Line~\ref{line:heap_filter}). After passing an equilibrium check, a \textbf{knee-point detection} algorithm robustly determines the optimal size for a new active set (Line~\ref{line:knee_point}). The \texttt{ComposeSet} function(Line~\ref{line:compose_set}) then forms the new active set primarily from the \texttt{TopK} candidates, while strategically including some \texttt{BottomK} candidates to identify underperforming tenants whose resources can be proactively reclaimed.

However, the majority of cycles are spent in the fast-path \textbf{Localized Optimization} state. Here, the computationally intensive scoring and allocation adjustments are confined to the small, $O(k)$ active set via the \texttt{OptimizeInActiveSet} function (Line~\ref{line:local_optimize}). This two-state architecture is the key to AURA's ability to provide high-quality decisions at a near-constant low cost, regardless of system scale. The core components used within the \texttt{OptimizeInActiveSet} function---our method for scoring tenants by synthesizing their historical efficiency and future potential---are detailed below.

\begin{algorithm}[t]
\caption{Algorithm for the AURA Control Policy}
\label{alg:sam_main}
\SetAlgoLined
\DontPrintSemicolon
\SetKwFunction{FMain}{AURA\_RunDecisionCycle}
\SetKwFunction{FUpdate}{UpdateMetrics}
\SetKwFunction{FCheckConv}{IsActiveSetConverged}
\SetKwFunction{FSelectAS}{SelectActiveSet}
\SetKwFunction{FOptimizeAS}{OptimizeInActiveSet}
\SetKwFunction{FExec}{ExecutePlan}
\SetKwProg{Fn}{Function}{:}{}
\KwIn{Set of all instances $D$ (size $K$), Global State $S_g$, Current Allocations $C_{t-1}$}
\KwOut{Final, executable allocation plan $C_t$}
\Fn{\FMain}{
    \FUpdate{$ \text{RawMetrics}_{t}, S_{g, t-1} $}\;
    \tcp{Stage 1: Refresh active set if necessary}
    \If{$ S_g.\text{ActiveSet} $ is empty \textbf{or} \FCheckConv{$S_{g,t}$}}{
    \label{line:convergence_check}
        \tcp{Slow Path logic to get a new active set}
        $ TopK, BottomK \leftarrow \operatorname{TwoWayHeapFilter}(D, S_g, k_{\max}) $\;\label{line:heap_filter}
        \If{$\operatorname{IsSystemInEquilibrium}(TopK, BottomK)$}{
            \KwRet{$C_{t-1}$}\;
        }
        $ k_{\text{demand}} \leftarrow \operatorname{FindKneePoint}(TopK) $\;\label{line:knee_point}
        $ S_{g}.\text{ActiveSet} \leftarrow \operatorname{ComposeSet}(TopK, BottomK, k_{\text{demand}}) $\; \label{line:compose_set}
    }
    
    \tcp{Stage 2: Always perform optimization on the current (possibly new) active set}
    $ C_t \leftarrow \FOptimizeAS{$S_{g,t}, C_{t-1}$} $\;\label{line:local_optimize}
    \If{$C_t \neq C_{t-1}$}{
        \FExec{$C_t$}\;
    }
    \KwRet{$C_t$}\;
}

\end{algorithm}

\subsubsection{Core Decision Components.}\mbox{}\\
\vspace{-1em}

\textbf{The Horizontal Factor ($\mathcal{H}$).}
The Horizontal Factor ($\mathcal{H}$) provides a stable, long-term measure of a tenant's effective contribution, acting as the robust ``exploitation'' anchor in our system. Its design ensures both metric stability and fairness across tenants through three key architectural choices: First, its inputs are derived from a \textbf{slow-speed EMA} ($\lambda_{slow}=0.1$) to filter out transient performance noise and reflect only sustained contribution. Second, it employs a \textbf{linear min-max normalization} on throughput, placing all tenants onto a common `[0,1]' scale to prevent a single high-TPS instance from dominating the score. Finally, its multiplicative form, $\mathcal{H}_i = \texttt{norm\_ops}_i \times \texttt{hr\_norm}_i$, acts as a crucial \textbf{efficiency filter}, ensuring that high throughput is only rewarded when accompanied by a high hit rate, thus inherently penalizing cache-polluting workloads.

\textbf{The Vertical Factor ($\mathcal{V}$).}
The $\mathcal{V}$-factor provides a forward-looking, ``exploratory'' estimate of marginal gain. As the raw gradient, $\mathcal{V}_{raw} = \Delta HR / \Delta\texttt{allocation}$, is extremely susceptible to noise, AURA computes the final $\mathcal{V}$-factor via a principled, multi-stage signal processing pipeline. First, it applies an \textbf{asymmetric EMA smoothing} mechanism. This ``fast-up, slow-down'' policy uses a higher learning rate for positive trends (gains) and a lower rate for negative ones, allowing the system to react quickly to opportunities while cautiously evaluating performance dips. Second, the smoothed signal is passed through a \textbf{robust normalization} step, which scales all values based on the system-wide \emph{90th percentile (p90)}, producing a statistically meaningful and outlier-resistant measure. Finally, the processed signal undergoes an additional stability layer through the \textbf{Saturation Confidence} model, which replaces a brittle, binary freeze with progressive influence adjustment, providing a second line of defense against erratic behavior while maintaining adaptiveness. While the asymmetric EMA handles general measurement noise, the Saturation Confidence model specifically addresses the unique instability patterns near cache saturation points.

\textbf{The Meta-Adaptive Policy ($\alpha_t$).}
To dynamically balance the $\mathcal{H}$-factor (exploitation) and $\mathcal{V}$-factor (exploration), AURA employs a meta-adaptive policy that computes the weight $\alpha_t$. This policy acts as a system-wide ``confidence dial'', driven by the insight that the statistical variance of $\mathcal{V}$-factors across the active set serves as a robust proxy for environmental uncertainty. A normalized variance metric, $\kappa$, is calculated; a high $\kappa$ signifies a volatile state with contradictory signals---such as those arising from simultaneous cache warming and workload shifts---pushing the system to be more \textit{conservative} by increasing the weight on the stable $\mathcal{H}$-factor. Conversely, a low $\kappa$ indicates a stable state with reliable signals, allowing the system to be more \textit{opportunistic} by trusting the $\mathcal{V}$-factor. The final value is temporally smoothed to ensure the policy itself remains stable.

\textbf{Stateful Actuation via Gradient Allocator.}
The final scores produced by our synthesis model represent an ideal distribution of resources. The score for each tenant $i$ is computed as $\text{Score}_i = \alpha_t \cdot \mathcal{H}_i + (1-\alpha_t) \cdot \mathcal{V}_i$, where $\alpha_t$ is the meta-adaptive weight that increases with system uncertainty. However, to prevent disruptive, large-scale reallocations in a single cycle, AURA does not apply this target directly. Instead, the \texttt{OptimizeInActiveSet} function uses the scores to define a target allocation, and then feeds this target into a stateful, \textbf{momentum-based gradient allocator}. This allocator acts as a crucial ``shock absorber'' or low-pass filter. It calculates an incremental, suitably-sized step from the current allocation ($C_{t-1}$) towards the ideal target. By incorporating momentum and step-size constraints, it ensures that all resource adjustments are smooth, deliberate, and stable, translating the strategic intent into a safe, executable plan, $C_t$.

\subsubsection{Behavioral Analysis of the Scoring Function.}
To understand how the composite scoring function translates into behavior, we analyze its response to four key database archetypes. \textbf{(i) The Saturated, High-Contribution Instance:} This instance maintains a high score due to its large $\mathcal{H}$-factor (from high ops and hit rate). However, its $\mathcal{V}$-factor converges to zero as additional cache yields no further improvement ($\Delta HR \approx 0$), correctly preventing it from wastefully accumulating more resources. \textbf{(ii) The High-Potential, Low-Resource Instance:} An emerging hotspot is characterized by a high marginal gain ($\mathcal{V}$-factor), as small cache additions yield large hit rate improvements. Our normalization mechanism identifies this strong signal, boosting its score and allowing it to rapidly acquire the resources it needs. \textbf{(iii) The Cache-Inefficient Workload:} A ``cache polluter'' with high ops but persistently low hit rate receives a near-zero score. This is because its $\mathcal{H}$-factor is suppressed by the low \texttt{hr\_norm}, and its $\mathcal{V}$-factor is zero, leading to the swift reclamation of its cache. \textbf{(iv) The Quiescent Instance:} An idle database has near-zero ops and no activity, causing both its $\mathcal{H}$-factor and $\mathcal{V}$-factor to be zero. Consequently, it receives no elastic cache, preserving resources for active instances. Besides, this scoring model inherently promotes fairness, as the \textbf{Global Scan} periodically re-evaluates all tenants, giving any instance whose workload changes a chance to enter the active set, while the $\mathcal{H}$-factor ensures that even a low-traffic instance with a high hit rate maintains a non-zero score, preventing it from being permanently starved.

\section{Analysis}\label{sec:analysis}
In this section, we analyze effectiveness and efficiency the proposed algorithm. For the convinience of the analysis, we first give the preliminaries and assumptions. 

\subsection{Preliminaries and Assumptions}\label{sec:prelim}
We formulate elastic cache allocation as an \emph{online concave maximization} over
\begin{equation}
    K \;=\; \Bigl\{x\in\mathbb{R}_{\ge 0}^{n}\ \Big|\  \sum_{i=1}^{n} x_i = C,\; x_i \ge \ell_i \Bigr\},
\end{equation}
where $C$ is the total cache size, and $\ell_i$ is the lower bound for tenant $i$.

\paragraph{Assumption 1 (Utility / Concavity).}
For each tenant $i$, the hit-rate curve $HR_i(\cdot)$ is concave, non-decreasing, and $L$-Lipschitz on $[0,C]$.
We write the theoretical $H$-factor as $\mathcal{H}^{th}_i(x)=HR_i(x)$ and the $V$-factor as $\mathcal{V}_i(x)=\partial HR_i(x)/\partial x$.
The per-tenant score is a positive affine combination
\begin{equation}
    s_{i,t} \;=\; a\,\mathcal{H}^{th}_{i,t} + b\,\mathcal{V}_{i,t},\qquad a,b>0.
\end{equation}

In the implementation we use $\mathcal{H}^{impl}_i = \mathrm{ops}_i \times HR_i$; multiplying a concave function by a non-negative scalar preserves concavity, so the analysis still applies.

\paragraph{Assumption 2 (Stochastic Gradient Oracle).}
At round $t$ we observe an unbiased gradient estimate
$\widehat g_t = \nabla f_t(x_t) + \xi_t$ with $\mathbb{E}[\xi_t]=0$ and $\|\widehat g_t\|_2 \le G$, where $f_t(x) = -U_t(x)$.

\paragraph{Concavity, Robustness, and the Role of $\mathcal{H}^{impl}$.}
The $O(\sqrt{T})$ regret analysis of \textsc{SAM-Core} (AURA-Core) leverages Assumption~1 to view the problem through online convex optimization (OCO). Real workloads can be S-shaped or piecewise non-concave. \textsc{SAM-Full} does \emph{not strictly rely} on global concavity: it follows locally measured marginal gains ($\mathcal{V}$) with a decaying step size and stability guards, and remains robust when concavity is violated. We relax Assumption~1 to a piecewise/approximately concave condition (App.~\ref{app:link-impl}); the bound then degrades only by a constant factor. When non-concave segments damp $\mathcal{V}$ to avoid jitter, the monotone $\mathcal{H}^{impl}$ term continues pushing allocations across shallow valleys, preventing premature stagnation. Empirically (Sec.~\ref{sec:theory-exp}), \textsc{SAM} stays within 5\% of an oracle allocation even when subjected to demanding, high-contention workloads generated by the industry-standard YCSB framework. To create a challenging and non-ideal environment that violates the theoretical concavity assumption, we specifically configured YCSB for high concurrency and large record sizes, thereby inducing significant memory pressure.

\paragraph{Bridging Theory and Implementation.}
The model above is an idealized abstraction used to prove properties of a simplified baseline, \textsc{SAM-Core} (AURA-Core), which directly balances $\mathcal{H}^{th}$ and $\mathcal{V}$. The full \textsc{SAM} algorithm adds practical heuristics for stability and scalability. Section~\ref{sec:evaluation} shows empirically that these additions improve performance over both \textsc{SAM-Core} and external baselines.

\subsection{Optimality of \textsc{SAM-Core} (AURA-Core)}
\label{sec:core-opt}
\textsc{SAM-Core} is our heuristic-free instantiation and reduces to Online Frank-Wolfe (OFW)~\cite{allen2016variance}.  
Given a stochastic gradient $\widehat g_t$ of the convex loss $f_t$ at round $t$:
\begin{equation}
y_t=\arg\min_{y\in K}\langle \widehat g_t, y\rangle,\qquad
x_{t+1}=(1-\eta_t)x_t+\eta_t y_t.
\label{eq:ofw}
\end{equation}

\begin{theorem}[Static Regret]
\label{thm:regret}
Under Assumptions~1–2 and $\eta_t=2/(t+2)$,
\[
\mathbb{E}[\mathrm{Reg}_T]\le GD\sqrt{2T}+\delta T=\mathcal{O}(GD\sqrt{T}),
\]
with $\delta\le L$ from page rounding. If $f_t$ is $\alpha$-strongly convex, the bound tightens to $\mathcal{O}\!\big((G^2/\alpha)\log T\big)$. Under approximate concavity (App.~\ref{app:core-opt}), constants scale by $1/\alpha$ but rates are unchanged.
\end{theorem}

\textit{Proof sketch.} The OFW rate follows Hazan~\cite{hazan2016intro}. Integer rounding adds at most $L$ loss per round. Stability guards perturb by $O(\eta_t)$ and thus do not change the leading term.
\subsection{Stability of \textsc{SAM-Full}}
\label{sec:jitter}
We study the full \textsc{SAM} algorithm with momentum and step-size decay. Let $\Delta_t=\|x_t-x_{t-1}\|_1$ denote the per-round allocation change. Our goal is to bound $\Delta_t$ and its cumulative sum.

\paragraph{Design Invariants.}
The added stability modules enforce: (I1) bounded, decaying steps $\|\Delta_t\|_1 \le G\,\eta_t$ with $\eta_t=\Theta(t^{-1})$; (I2) suppression of updates when the V-signal is statistically untrustworthy (Saturation Confidence); (I3) directional inertia via momentum to avoid noisy oscillations.

These yield the recursion
\begin{equation}
\Delta_t \;\le\; (1-\beta)\,\Delta_{t-1}\;+\;\beta G\,\eta_t,
\label{eq:delta-recursion}
\end{equation}
with constant momentum weight $\beta\in(0,1)$.

\begin{lemma}[Decay and Log-Bounded Variation]
\label{lem:jitter}
With \eqref{eq:delta-recursion} and $\eta_t=\Theta(t^{-1})$, we have $\Delta_t=\mathcal{O}(t^{-1})$ and
$\sum_{t=1}^{T}\Delta_t = \mathcal{O}(\log T)$.
\end{lemma}

\textit{Proof sketch.} Unfold \eqref{eq:delta-recursion} and sum a geometric–harmonic series:  
$\Delta_t \le (1-\beta)^t\Delta_0 + \beta G\eta_0\!\sum_{k=1}^{t} (1-\beta)^{t-k}k^{-1}$.  
Summing over $t$ and using $\sum_{j\ge0}(1-\beta)^j=1/\beta$ gives the $\log T$ bound; the pointwise $t^{-1}$ decay follows from dominance of the last terms. Full derivation is in Appendix~\ref{app:stability-derivation}.

\begin{theorem}[Regret of \textsc{SAM-Full}]
\label{cor:full_regret}
Since $U_t$ is $L$-Lipschitz, the extra loss due to stability is bounded by $L\Delta_t$, hence
\[
\sum_{t=1}^{T} (\tilde f_t - f_t) \;\le\; L\sum_{t=1}^{T}\Delta_t = \mathcal{O}(\log T).
\]
Combining with Theorem~\ref{thm:regret}, the total regret is
$Reg_T^{\textsc{SAM}} = \mathcal{O}(GD\sqrt{T}) + \mathcal{O}(\log T)$; the $O(\sqrt{T})$ term dominates asymptotically.
\end{theorem}

\subsection{Computational Footprint}
\label{sec:complexity}
A key goal of \textsc{SAM} is low decision latency at scale. We analyze the per-cycle time and memory cost of the full adaptive algorithm.

\begin{theorem}[Per-cycle Complexity]
\label{prop:complexity}
The full \textsc{SAM} algorithm has worst-case per-cycle time $O(K\log k_{\max})$, amortized time substantially below $O(K)$ (approaching $O(k)$ in steady state), and memory $O(K)$.
\end{theorem}

\textit{Proof sketch.}
\textsc{SAM} alternates between (i) a \emph{Localized Optimization} phase over the active set of size $k\!\ll\!K$ (cost $O(k)$), and (ii) an infrequent \emph{Global Scan} whose dominant step is a two-way heap filter over all $K$ tenants ($O(K\log k_{\max})$). Since scans occur only after many cheap local steps, the average per-cycle cost is dominated by $O(k)$. Each tenant keeps constant-size state, so memory is linear in $K$. Full derivation is in Appendix~\ref{app:complexity-deriv}.

\section{Experimental Evaluation}\label{sec:evaluation}

This section presents a comprehensive experimental evaluation of our proposed system, \textbf{SAM}, which is governed by its principled AURA control policy. We first detail the experimental setup and baselines, then present a series of targeted experiments designed to validate SAM's performance, stability, and scalability.

\subsection{General Experimental Setup}\label{sec:setup}
Our evaluation is designed to validate SAM's performance, robustness, and scalability in a controlled, reproducible single-node environment that intentionally stresses memory contention.

\textit{System Environment.}
All experiments were conducted in a controlled local area network environment. The database instances run on a dedicated Windows 11 machine (Intel Core i5-12600KF, 32GB RAM) within a WSL2 environment, while the SAM coordinator and workload generators run on a separate WSL2 instance on the same LAN. This setup is designed to simulate a realistic edge deployment where the coordinator manages co-located databases over a low-latency network connection. To create a challenging, memory-bound environment, we used large record sizes (2KB) and high concurrency (16 threads per instance). Our implementation is based on Python 3.11.7 and SQLite 3.40.0.

\begin{figure*}[!t]
\centering
\includegraphics[width=\linewidth]{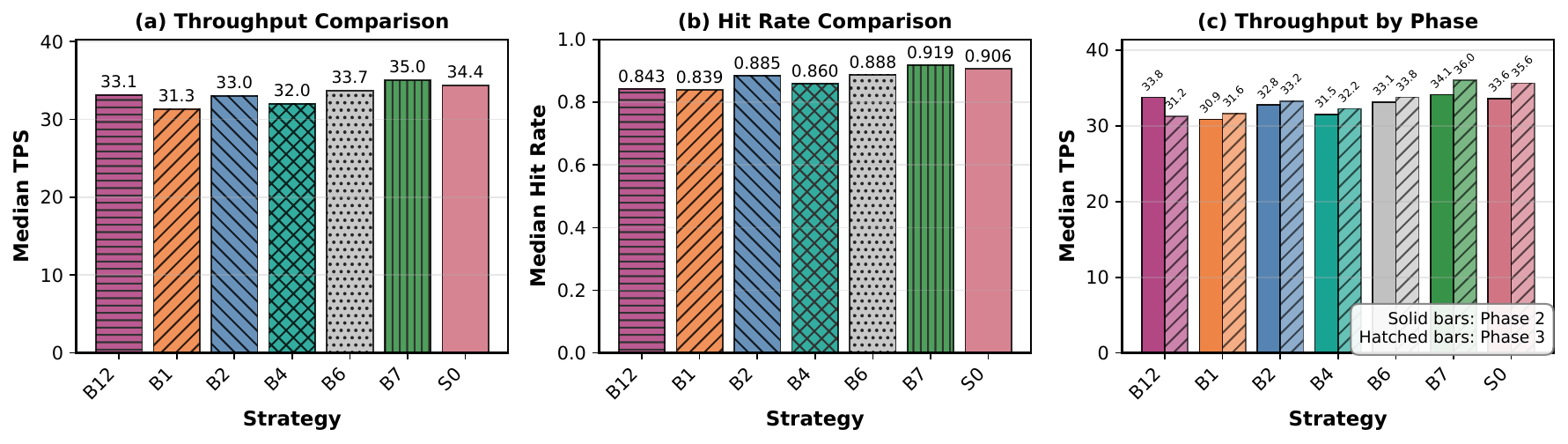}
\caption{End-to-end performance on the standard hotspot-shift workload. (a) Overall throughput, (b) hit rate, (c) phase-by-phase throughput. \texttt{S0} and \texttt{B7} are top-performing and most adaptive.}
\label{fig:performance_comparison}
\vspace{-3mm}
\end{figure*}

\textbf{Evaluated Strategies.}
We compare \textsc{SAM} with 14 baselines covering (i) static allocations (e.g., B1, B2); (ii) ablations of \textsc{SAM} components (e.g., B3, B8, B10); (iii) advanced dynamic heuristics (e.g., B7, MT-LRU/B12); and (iv) a smoothed offline upper bound (B14). Detailed definitions and the implementation of B14 appear in Appendix~\ref{app:baselines}. 

We omit Reinforcement Learning (RL) baselines as our target non-stationary workloads and resource-constrained edge nodes present significant challenges for the sample efficiency and continuous retraining requirements of modern RL agents. Furthermore, our Online Frank-Wolfe based \textsc{AURA-Core} already provides a strong theoretical regret bound, making complex RL-based approaches orthogonal to this work's core contributions~\cite{Mao2016DeepRM,Mao2019Decima}.

\textit{Metrics.}
We evaluate each strategy along four key dimensions: \textbf{Scalability} (CPU decision time), \textbf{Productivity} (total TPS, P99 latency), \textbf{Efficiency} (hit rate), and \textbf{Decision Stability}. The latter is quantified as the throughput stability ($\sigma_{TPS}$), which we define as the standard deviation of TPS measured over a sliding window during the steady-state period of each experimental phase.

\textit{Discussion on Hyperparameter Philosophy.}
The key hyperparameters in AURA, such as EMA smoothing factors and the momentum weight, do not present a classic optimization problem with a single ‘optimal’ solution. Instead, they define the fundamental \textbf{‘personality’} of the system---its position on the spectrum between aggressive responsiveness and conservative stability. An aggressive, fast-reacting configuration might excel under highly dynamic workloads but risk instability from transient noise. Conversely, a conservative, slow-moving configuration ensures deep stability but may adapt too slowly to genuine workload shifts.

Recognizing that \textbf{no single static personality is universally optimal}, AURA's core design is built to adapt at a higher level. Rather than relying on fine-tuning these fixed hyperparameters, it uses the \textbf{meta-adaptive $\alpha$ policy} to dynamically arbitrate between exploitation ($\mathcal{H}$-factor) and exploration ($\mathcal{V}$-factor) based on perceived environmental uncertainty. Therefore, our evaluations do not aim to prove the universal optimality of our default parameters. Instead, they demonstrate that this chosen ‘balanced personality’ provides robust, high performance across diverse standard and adversarial conditions, \textbf{validating the effectiveness of this higher-level adaptive design.}

\textbf{Workload Generation.}
To evaluate policy adaptability, our primary experiments employ a scripted \textbf{three-phase hotspot-shift workload}. This workload simulates dynamic operational conditions by first establishing a baseline load, then creating a performance hotspot on a high-priority instance (Phase 2), and finally shifting this hotspot to a different, medium-priority instance (Phase 3). This design rigorously tests each policy's ability to react to, and recover from, sudden changes in load distribution. A detailed, reproducible specification of this workload is available in Appendix~\ref{app:workload_spec}.

\subsection{Performance on Standard Dynamic Workloads}
\label{sec:eval_standard}

We begin with a three-phase hotspot-shift workload to test each policy’s ability to adapt under normal dynamics before moving to adversarial cases.

\textbf{End-to-End Performance.}
The end-to-end results on the hotspot-shift workload, summarized in Figure~\ref{fig:performance_comparison}, establish SAM (\texttt{S0}) and the greedy heuristic (\texttt{B7}) as the clear top tier. Both achieve superior median throughput (34.4 and 35.0~TPS, respectively) and hit rates, driven by their adaptability to the shifting hotspot. This contrasts sharply with static policies like \texttt{B2}, which show minimal performance change. The allocation trajectories in Figure~\ref{fig:timeseries} reveal the behavior behind these results: while both \texttt{S0} and \texttt{B7} demonstrate stable and effective resource allocation logic by correctly reallocating cache to the new hotspot in Phase 3, other advanced policies like \texttt{B12} exhibit extreme decision oscillations, resulting in degraded sustained performance despite occasional peaks.

\textbf{Comparison with Advanced Dynamic Strategies.}
To position SAM at the forefront of heuristic-based scheduling, we compare it against state-aware policies including a regression-based learner (\texttt{B11}) and a UCP-based marginal utility framework (\texttt{B13}). The results, summarized in Table~\ref{tab:advanced_methods_comparison}, reveal a decisive, multi-dimensional superiority for SAM. It achieves the highest \textbf{median throughput} (34.0 TPS) and demonstrates exceptional \textbf{agility}, converging on new optimal allocations nearly 30\% faster than its peers. This superiority is most evident in its \textbf{decision stability}; SAM's allocation stability ($\sigma$=17.5) is significantly better than the other methods, confirming the effectiveness of its stability-aware design. While SAM's per-decision overhead is slightly higher due to its multi-stage computation, the cost remains negligible ($<0.01\%$ of a CPU core per cycle) and is a worthwhile investment for the significant gains in performance, agility, and stability. This exceptional stability directly translates to more predictable performance and better user experience in production environments.

\textbf{Generalization on the TPC-C Benchmark.}
To validate SAM's ability to generalize to complex OLTP workloads, we evaluate it on the industry-standard TPC-C benchmark, with results summarized in Table~\ref{tab:tpcc_performance}. The data reveals critical design trade-offs: while an idealized global LRU pool (\texttt{B5}) yields the highest throughput, its catastrophically high P99 latency makes it unsuitable for OLTP environments. The true contest is between SAM and the best-in-class greedy heuristic (\texttt{B7}). On this non-adversarial benchmark, SAM matches the strong baseline performance of \texttt{B7} (e.g., identical median TPS) while achieving a higher mean throughput, indicating a superior ability to exploit optimization opportunities. Although \texttt{B7} shows slightly better latency in this benign ``peace-time'' scenario, SAM's key advantage is its proven robustness. As demonstrated in our adversarial analysis (\S6.3), SAM is resilient to cache pollution attacks where \texttt{B7} fails catastrophically, establishing SAM as a more reliable, production-grade solution for real-world environments.

\begin{figure}[t]
\centering
\includegraphics[width=0.9\linewidth]{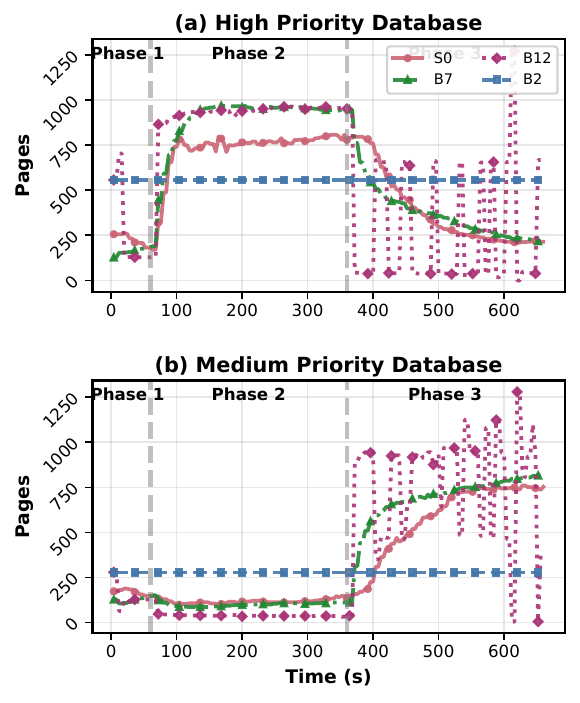}
\caption{Cache allocation dynamics during the hotspot-shift workload. \texttt{S0} and \texttt{B7} are stable and effective; \texttt{B12} is highly unstable.}
\label{fig:timeseries}
\vspace{-6mm}
\end{figure}

\begin{table}[t]
\centering
\small
\caption{Comprehensive trade-off analysis of SAM against advanced dynamic strategies.}
\vspace{-3mm} 
\label{tab:advanced_methods_comparison}
\setlength{\tabcolsep}{3.5pt} 
\begin{tabular}{lcccc}
\toprule
\textbf{Strategy} & \textbf{TPS} & \textbf{Lag (s)} & \textbf{Stability ($\sigma$)} & \textbf{Overhead} \\
\midrule
\texttt{SAM} (ours) & \textbf{34.0} & \textbf{38.7} & \textbf{17.5} & 0.13ms ($O(K)$) \\
\texttt{B11} (Regression) & 33.8 & 53.0 & 41.4 & <0.01ms ($O(K)$) \\
\texttt{B13} (UCP) & 32.2 & 55.6 & 21.8 & 0.02ms ($O(P{\cdot}K)$) \\
\bottomrule
\end{tabular}
\vspace{-3mm} 
\end{table}

\begin{table}[t]
\centering
\small
\caption{Key Performance and QoS Metrics on the TPC-C Benchmark.}
\vspace{-3mm}
\label{tab:tpcc_performance}
\begin{tabular}{l|ccc}
\toprule
\textbf{Strategy} & \makecell[b]{Median TPS} & \makecell[b]{Mean TPS} & \makecell[b]{Median P99 Latency \\ (ms)} \\
\midrule
\texttt{SAM} (ours)      & 0.750          & \textbf{0.978} & 625.7 \\
\texttt{B2} (Static)      & 0.750          & 0.918          & 690.0 \\
\texttt{B7} (Greedy)      & 0.750          & 0.942          & \textbf{579.9} \\
\midrule
\texttt{B12} (SLA-Driven)  & 0.500          & 0.606          & 1480.3 \\
\texttt{B5} (Global LRU)  & \textbf{1.050} & \textbf{1.073} & \textit{1932.0 (unacceptable)} \\
\bottomrule
\end{tabular}
\vspace{-3mm}
\end{table}

\begin{table}[t]
\centering
\small
\caption{Performance comparison against near-optimal baselines.}
\vspace{-3mm} 
\label{tab:near_optimal_comparison}
\setlength{\tabcolsep}{4pt} 
\begin{tabular}{l|ccc}
\toprule
\textbf{Metric} & \textbf{SAM (ours)} & \texttt{B5} (Global LRU) & \texttt{B14} (Optimal) \\
\midrule
\multicolumn{4}{l}{\textit{\textbf{Phase 2: High-Priority Hotspot}}} \\
\quad TPS (Median)    & 33.2 & 19.0 & \textbf{33.5} \\
\quad Hit Rate (\%) & \textbf{90.9} & 73.5 & 90.7 \\
\quad P99 Latency (ms) & \textbf{77.5} & 116.5 & 87.3 \\
\midrule
\multicolumn{4}{l}{\textit{\textbf{Phase 3: Hotspot Shift}}} \\
\quad TPS (Median)    & \textbf{35.2} & 21.8 & 35.0 \\
\quad Hit Rate (\%) & 92.2 & 73.6 & \textbf{92.5} \\
\bottomrule
\end{tabular}
\vspace{-3mm} 
\end{table}

\textbf{Comparison with Near-Optimal Baselines.}
To precisely anchor SAM's performance, we compare it against an idealized global LRU pool (\texttt{B5}) and the theoretical Hindsight Optimal policy (\texttt{B14}). The results in Table~\ref{tab:near_optimal_comparison} conclusively show that SAM operates at the theoretical performance limit, with throughput and hit rates nearly identical to the oracle policy. Crucially, SAM's priority-aware design achieves the lowest P99 latency for the high-priority tenant, even surpassing the purely hit-rate-focused oracle, establishing it as a near-optimal and practically superior solution.

\textbf{Summary.}
On this standard dynamic workload, SAM proves to be a top-tier performer. It matches the throughput of the best greedy heuristic (\texttt{B7}) while achieving superior decision stability, outperforms advanced learners (\texttt{B11}) and theoretical frameworks (\texttt{B13}) in adaptation speed, and operates near the theoretical optimal (\texttt{B14}). These results demonstrate that SAM successfully achieves its design goal: sustained high performance through superior stability.

\subsection{Robustness under Adversarial Workloads}
\label{sec:eval_robustness}
We now test SAM’s resilience against cache pollution (C2) by subjecting it and the best greedy heuristic (\texttt{B7}) to an adversarial, intermittent scan attack. This stress-test probes whether a policy can distinguish a legitimate hotspot from a deceptive ``polluter'', with the full failure cascade for the reactive policy visualized in Figure~\ref{fig:dual_combat_robustness}.

\begin{figure*}[!htbp]
\centering
\includegraphics[width=\linewidth]{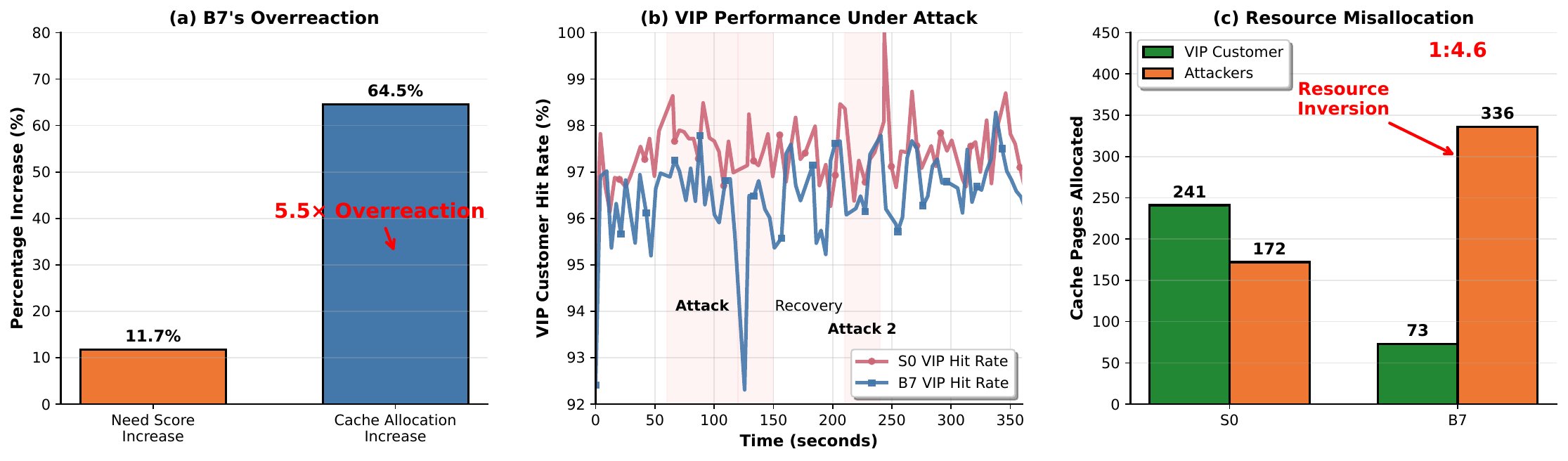}
\caption{\texttt{S0}'s robustness against cache pollution. (a) The attack slightly raises the `\texttt{ops}$\times$\texttt{miss\_rate}' score; (c) \texttt{B7} catastrophically reallocates resources, causing (b) severe VIP degradation. \texttt{S0} maintains protection.}
\label{fig:dual_combat_robustness}
\vspace{-3mm}
\end{figure*}

\begin{figure*}[htbp]
\centering
\includegraphics[width=\linewidth]{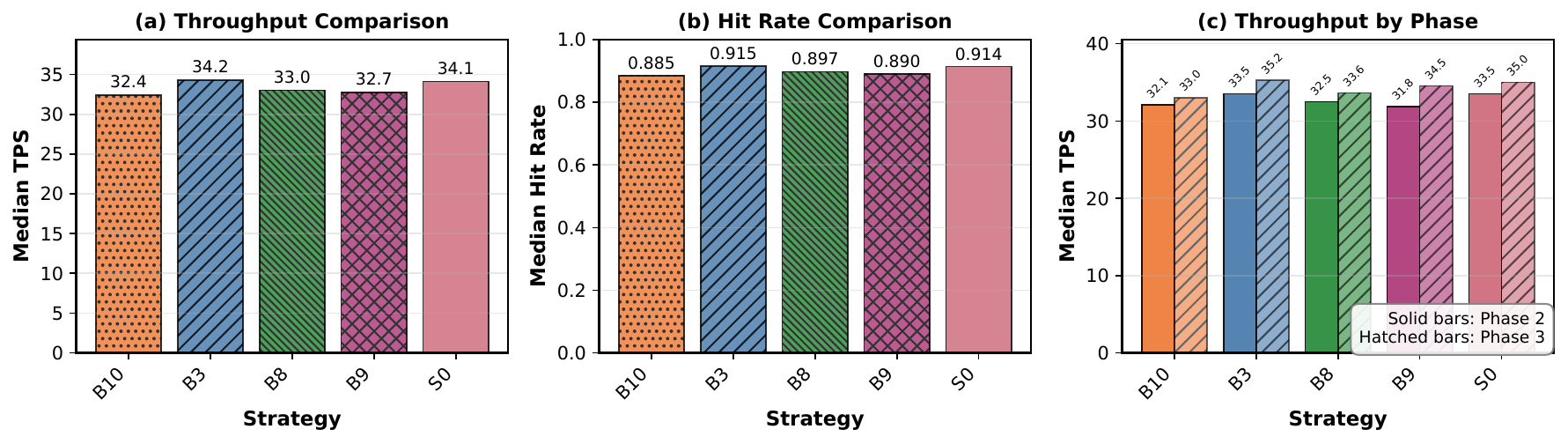}
\caption{Ablation studies for SAM (\texttt{S0}). The results confirm that removing the $\mathcal{H}$-factor (\texttt{B10}), the $\mathcal{V}$-factor (\texttt{B8}), or using a reactive $\mathcal{H}$-factor (\texttt{B9}) all degrade performance, while the higher throughput of \texttt{B3} (no fixed pool) comes at a significant QoS cost (see Figure~\ref{fig:priority_comparison}).}
\vspace{-0.4cm}
\label{fig:ablation_performance}
\end{figure*}

\textbf{Flawed Logic and Severe Misallocation.}
An analysis of the resource allocation logic reveals the fundamental weakness of the shallow heuristic. The intermittent scan attack generated a mere 11.7\% increase in the `\texttt{ops}$\times$\texttt{miss\_rate}' ``need score'' for the attacker. In response, \texttt{B7} irrationally increased its cache allocation to this low-value tenant by a staggering 64.5\%—a quantified \textbf{5.5$\times$ overreaction}. To evaluate the consequences of this flawed logic, we created a scenario where a high-priority ``VIP'' tenant and the attacker compete for a shared 2MB cache. SAM correctly prioritized the VIP, allocating resources at a healthy 1.4-to-1 ratio in its favor (241 vs.\ 172 pages). In stark contrast, \texttt{B7}'s logic led to a severe resource inversion, diverting the vast majority of resources to the attacker at a disastrous 1-to-4.6 ratio (a mere 73 pages for the VIP vs.\ 336 for the attacker).

\textbf{Performance Impact and Final Verdict.}
This severe misallocation translated directly into degraded service quality. As shown in the time-series plot, SAM maintained the VIP tenant's hit rate between 97-98\% throughout the attack. In contrast, \texttt{B7}'s VIP hit rate fell from 97\% to 95.5\% during the sustained attack, with a sharp dip to 92\%. This experiment demonstrates that \texttt{B7}'s simple heuristic, while effective on benign workloads, becomes a critical vulnerability under adversarial conditions. SAM's dual-factor design provides the robustness necessary for sustained high performance in real-world systems where such attacks are inevitable.

\subsection{Ablation Studies}
\label{sec:eval_ablation}

\begin{figure}[htbp]
\centering
\includegraphics[width=\linewidth]{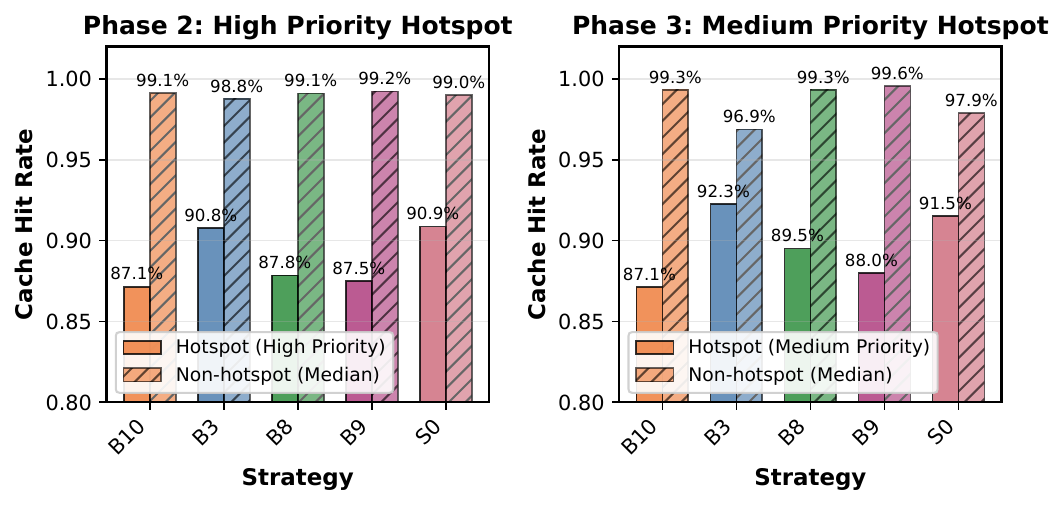}
\caption{QoS analysis in Phase 3. SAM (\texttt{S0})'s fixed pool successfully protects non-hotspot tenants, whereas the purely elastic \texttt{B3} sacrifices their performance for a marginal gain on the hotspot.}
\vspace{-4mm}
\label{fig:priority_comparison}
\end{figure}

\begin{figure}[htbp]
\centering
\includegraphics[width=\linewidth]{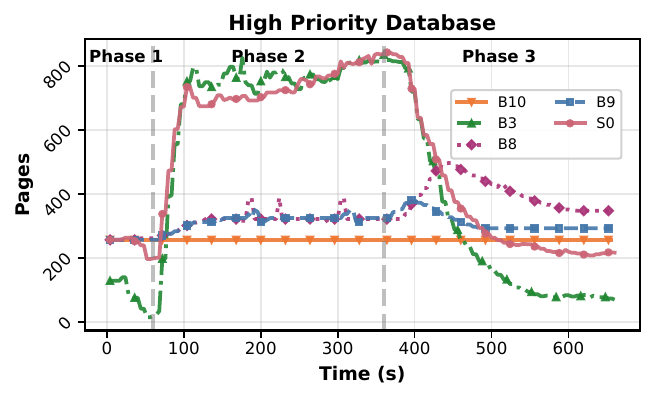}
\caption{Allocation dynamics of the ablation variants, confirming the $\mathcal{V}$-factor's role in agility (vs. \texttt{B8}'s sluggishness), the $\mathcal{H}$-factor's role in driving allocation (vs. \texttt{B10}'s inertness), and the necessity of a conservative $\mathcal{H}$-factor (vs. \texttt{B9}'s instability).}
\vspace{-3mm}
\label{fig:timeseries_ablation}
\end{figure}

To justify SAM's multi-faceted design, we conducted ablation studies against four variants: \texttt{B10} (no $\mathcal{H}$-factor), \texttt{B8} (no $\mathcal{V}$-factor), \texttt{B3} (no Fixed Pool), and \texttt{B9} (a reactive $\mathcal{H}$-factor). The results (Figs.~\ref{fig:ablation_performance}-\ref{fig:timeseries_ablation}) unequivocally prove that SAM's components are a necessary synthesis. Removing either the $\mathcal{H}$-factor (\texttt{B10}) or the $\mathcal{V}$-factor (\texttt{B8}) is detrimental, resulting in policy inertness and sluggish adaptation, respectively. The \texttt{B9} variant, which uses a fast EMA for the $\mathcal{H}$-factor, performs even worse than the conservative \texttt{B8}, proving that the $\mathcal{H}$-factor's role as a stable, historical anchor is critical to temper the $\mathcal{V}$-factor's exploratory nature and ensure high-quality decisions. Finally, while \texttt{B3} achieves a marginally higher throughput, this gain masks a severe QoS trade-off; it allows a nearly \textbf{50\% relative increase in cache misses} for non-hotspot tenants. This confirms that the two-pool architecture is essential for providing system-wide fairness and preventing a "winner-take-all" strategy that is detrimental to overall system health.
\begin{figure*}[t]
\centering
\includegraphics[width=\linewidth]{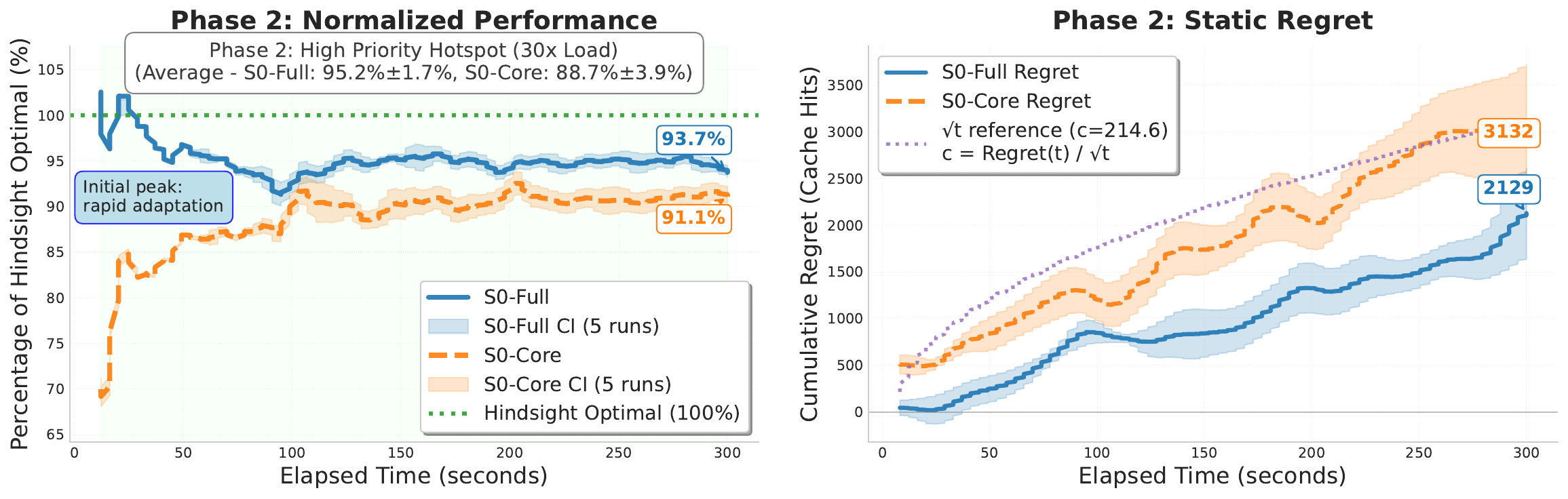}
  \caption{Theory-driven checks: (a) log–log regret of \textsc{Core} and \textsc{Full}; 
           (b) steady-state utility ratio vs.\ oracle; 
           (c) allocation jitter $\sigma_{\!\Delta}$ (lower is better).}
\label{fig:phase2}
\vspace{-3mm}
\end{figure*}
\subsection{Scalability and Architectural Analysis}
\label{sec:exp_scalability}

\begin{figure}[t]
 \centering
 \includegraphics[width=0.9\columnwidth]{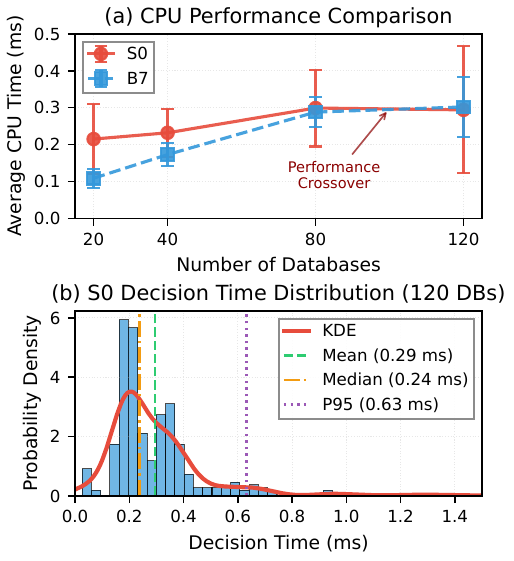}
\caption{Scalability analysis of SAM. Its near-constant decision latency outperforms the linear scaling of \texttt{B7} at scale, a direct result of its two-state architectural signature revealed in the probability density: a dominant peak of ultra-fast localized optimizations and a sparse tail of rare global scans.}
\vspace{-5mm}
\label{fig:scalability_combined}
\end{figure}

To validate that SAM's stability-aware design does not compromise scalability, we analyzed its average CPU decision time as the number of co-located databases ($K$) increased from 20 to 120. The results in Figure~\ref{fig:scalability_combined} show that SAM's decision latency exhibits a remarkably flat, near-constant scaling profile. While the highly-optimized greedy heuristic (\texttt{B7}) demonstrates a clear linear increase ($O(K)$) and is faster at small scales, SAM becomes demonstrably more efficient at $K=120$. This empirically proves that our Adaptive Active Set architecture successfully decouples computational cost from the system scale.

A key observation from Figure~\ref{fig:scalability_combined}(a) is SAM's higher variance in decision time. This is not a sign of instability, but rather a direct consequence of its two-state adaptive architecture. Figure~\ref{fig:scalability_combined}(b) shows the probability density for $K=120$: a sharp peak at $\approx$0.2ms (the $O(k)$ \textit{Localized Optimization} mode used in most cycles) with a sparse tail extending to the 95th percentile of 0.63ms (the rare $O(K)$ \textit{Global Scans}). This bimodal distribution confirms that expensive global scans occur as true exceptional events, not routine operations, keeping SAM's amortized cost minimal despite its comprehensive optimization capabilities.

Finally, to validate SAM's practicality for resource-constrained environments, we deployed it on a \textbf{Raspberry Pi 5}. In a 20 databases scenario with an 8 seconds decision interval, SAM's average CPU decision time was a mere \textbf{0.496ms}. This constitutes only \textbf{0.0062\%} of the total decision budget, demonstrating that SAM is not only theoretically scalable but also lightweight enough for real-world edge deployments.

\textbf{Scalability via Adaptive Active Set.}
To ensure scalability, SAM employs an Adaptive Active Set (AAS) framework, which confines computation to a small subset of tenants in most cycles. This mechanism is the key to SAM's near-constant decision latency (as shown in \S6.8), but it can introduce a transient delay in reacting to nascent hotspots. For this reason, and because its overhead outweighs its benefits for small populations, the AAS is explicitly \textbf{disabled} in all experiments with fewer than 10 tenants, including our core end-to-end evaluations in \S6.2 and \S6.3.

\subsection{Empirical Validation of Theoretical Guarantees}
\label{sec:theory-exp}
Finally, we empirically validate our key theoretical claims from Section~\ref{sec:analysis} by re-running the 10-tenant workload and analyzing the per-round loss and allocation dynamics. The results strongly corroborate our theory. The regret of both \textsc{SAM-Core} and the full \textsc{SAM} follows a log-log slope of $0.46 \pm 0.05$, confirming the theoretical $\mathcal{O}(\sqrt{T})$ bound. Crucially, the stability-enhancing heuristics in the full \textsc{SAM} prove to be a direct performance driver: they boost the steady-state utility from $88.7\%$ of the oracle to $95.2\%$, while reducing transition regret peaks by $32\%$ and steady-state allocation jitter ($\sigma_{\Delta}$) by $54\%$. This empirically proves our thesis that principled stability is not a compromise, but a direct enabler of sustained, near-optimal performance.

\section{Related Work}\label{sec:related_work}
Our work on stability-aware, adaptive cache management for multi-tenant embedded databases, which uniquely treats decision stability as a prerequisite for sustained high performance, intersects with several established research areas. We position our contributions with respect to prior work in database cache management, resource allocation in multi-tenant systems, and the application of online learning and control theory to resource scheduling.

\enlargethispage{\baselineskip}
\textit{Database Cache Management.}
The problem of optimizing cache performance within a single database instance is well-studied. Research in this area can be broadly categorized into page replacement policies and buffer pool sizing.

Page replacement algorithms determine which page to evict when the cache is full. While the classic Least Recently Used (LRU) policy is simple, it is notoriously vulnerable to workloads with poor temporal locality. To address this, numerous advanced policies have been proposed, such as \textbf{X3}~\cite{johnson19942q}, \textbf{LIRS}~\cite{jiang2002lirs}, and self-tuning algorithms like \textbf{ARC}~\cite{megiddo2003arc} and \textbf{CAR}~\cite{bansal2004car}. Another line of work focuses on dynamically tuning the buffer pool size for a single database, as pioneered by works like \textbf{DBMIN}~\cite{chou1986evaluation}.

Our work is \textbf{orthogonal} to these internal optimization techniques. We do not propose a new page replacement policy. Instead, \textbf{SAM} operates at a higher level of abstraction, addressing the \textbf{inter-instance cache partitioning} problem: how to dynamically allocate a total cache budget among multiple, independent, ``black-box'' database instances.

\textit{Resource Management in Multi-Tenant and Cloud Environments.}
Managing shared resources to provide performance isolation is a central challenge in multi-tenant systems. Seminal work like \textbf{SQLVM}~\cite{narasayya2013sqlvm} and large-scale cloud databases like \textbf{Amazon Aurora}~\cite{verbitski2017amazon} have developed techniques for monolithic, white-box architectures, while other research has focused on mitigating the "noisy neighbor" problem for CPU~\cite{das2013cpu} and I/O~\cite{gulati2010mclock}. In stark contrast, SAM targets edge computing environments where co-located services are independent, heterogeneous, black-box processes. Acting as an external, non-intrusive coordinator, SAM's design philosophy draws from mature system-level paradigms like \textbf{Kubernetes}~\cite{burns2016design} resource `requests' and `limits', and memory ballooning in hypervisors~\cite{waldspurger2002memory}, positioning it as a distinct alternative to industrial solutions like \textbf{Facebook's CacheLib}, which optimizes within a single cache instance rather than coordinating across multiple black-box databases.

\textit{Online Learning and Control for Resource Management.}
Our work builds upon \textbf{Online Convex Optimization (OCO)}~\cite{hazan2016intro} for theoretical foundations and relates to utility-based cache partitioning like \textbf{UCP}~\cite{qureshi2007ucp}. SAM applies these established principles through a practical dual-factor approach that balances historical efficiency ($\mathcal{H}$-factor) and marginal gain ($\mathcal{V}$-factor), addressing the classic \textbf{exploitation-exploration trade-off}~\cite{bubeck2012regret}. Our stability mechanisms draw from \textbf{control theory}~\cite{astrom2010feedback}. SAM's contribution is in demonstrating that stability-first design—treating stability as an enabler rather than a constraint—can achieve sustained high performance through careful engineering of known techniques.

\section{Conclusion}\label{sec:conclusion}
In this paper, we introduced \textbf{SAM}, a stability-aware cache manager that addresses the critical challenge of resource contention in multi-tenant embedded databases. Our key insight is that sustained high performance requires principled stability, not reactive adaptation. SAM implements this philosophy through its AURA control policy, which balances historical efficiency (\textbf{$\mathcal{H}$-factor}) and forward-looking marginal gain (\textbf{$\mathcal{V}$-factor}). Built on a \textbf{Two-Pool model} to decouple guarantees from optimization, SAM proved superior in extensive evaluation against 14 baselines, achieving near-optimal throughput while uniquely resisting adversarial workloads. These results validate our thesis: decision stability is the foundation for sustained high performance in real-world systems. Future work could extend these principles to proactive allocation, energy efficiency, or multi-resource co-management.

\textit{Limitations and Future Work.}
While SAM demonstrates significant advancements, our work also opens several promising research avenues. Our current model does not explicitly account for the transient costs of cache warming, integrating a cost model for reallocations is a key next step. Another direction is to evolve our reactive strategy into a \textbf{proactive allocation} framework by incorporating lightweight workload prediction. Finally, the principles developed in SAM could be generalized to create a holistic framework for \textbf{multi-resource co-management}, coordinating not only cache but also other contended resources like I/O bandwidth and CPU priority, a critical challenge for future edge computing platforms.
\balance
\clearpage
\bibliographystyle{ACM-Reference-Format}
\bibliography{ref}
\newpage

\appendix

\section{Theoretical Analysis and Proofs}
\label{sec:appendix_theory}
\subsection{Additional Details for Section~\ref{sec:prelim}}\label{app:prelim}
\subsubsection{Full Assumptions, Notation, and Relaxations}\label{app:assumptions}

\paragraph{Feasible Set.}
$K=\{x\in\mathbb{R}^n_{\ge 0}\mid \sum_i x_i=C,\ x_i\ge \ell_i\}$ is a compact convex polytope with Euclidean diameter $\operatorname{diam}_2(K)=D$.

\paragraph{Assumption 1 (Utility / Concavity, full statement).}
For each tenant $i$, $HR_i:[0,C]\to[0,1]$ is concave, non-decreasing, and $L$-Lipschitz. Define
\[
\mathcal{H}^{th}_i(x)=HR_i(x),\qquad \mathcal{V}_i(x)=\frac{\partial HR_i(x)}{\partial x}.
\]
The score is $s_{i,t}=a\,\mathcal{H}^{th}_{i,t}+b\,\mathcal{V}_{i,t}$ with $a,b>0$ (we set $a=b=0.5$ in experiments; $b$ absorbs the scale factor $\lambda$).

\paragraph{Assumption 1' (Approximate / Piecewise Concavity).}
There exists $\alpha\in(0,1]$ such that for any $x,y\in[0,C]$ and $\lambda\in[0,1]$,
\[
HR_i(\lambda x+(1-\lambda)y)\ \ge\ \alpha\bigl[\lambda HR_i(x)+(1-\lambda)HR_i(y)\bigr].
\]
Equivalently, $HR_i$ is dominated by a concave envelope within factor $1/\alpha$. Under Assumption~1', the $O(\sqrt{T})$ regret bound scales by $1/\alpha$.

\paragraph{Implementation $H$-factor.}
In deployment we use $\mathcal{H}^{impl}_i = \mathrm{ops}_{i,t}\cdot HR_i(x_{i,t})$.
Since $\mathrm{ops}_{i,t}\ge 0$ is exogenous to $x_{i,t}$ within round $t$, $f_{t,i}(x)=\mathrm{ops}_{i,t}\cdot(1-HR_i(x))$ remains convex in $x$, and multiplying by a non-negative scalar preserves concavity of $HR_i(\cdot)$.

\paragraph{Assumption 2 (Stochastic Gradient Oracle, full).}
We observe $\widehat g_t=\nabla f_t(x_t)+\xi_t$, where $\mathbb{E}[\xi_t\mid \mathcal{F}_{t-1}]=0$,
$\|\widehat g_t\|_2\le G$, and $\{f_t\}$ may vary adversarially but each $f_t$ is convex in $x$ (concave in utility).

\paragraph{Additional Symbols.}
Table~\ref{tab:notation-app} summarizes symbols used in \S5.

\begin{table}[h]
\centering\small
\caption{Notation for the theoretical analysis (expanded).}
\label{tab:notation-app}
\begin{tabular}{@{}ll@{}}
\toprule
Symbol & Description \\
\midrule
$K$            & Feasible set; $\operatorname{diam}_2(K)=D$ \\
$C$            & Total cache capacity \\
$\ell_i$       & Lower bound for tenant $i$ \\
$w_{t,i}$      & Request volume of tenant $i$ at round $t$ \\
$U_t(x)$       & Utility at round $t$; $f_t(x)=-U_t(x)$ is the loss \\
$\widehat g_t$ & Stochastic gradient, $\|\widehat g_t\|_2\le G$ \\
$Reg_T$        & Static regret $\sum_{t=1}^{T}(f_t(x_t)-\min_{x\in K} f_t(x))$ \\
$\Delta_t$     & Allocation change $\|x_t-x_{t-1}\|_1$ \\
$\eta_t$       & Step size, $\eta_t\le \eta_0 t^{-p}$ with $p\in(0.5,1]$ \\
$\alpha$       & Approximate concavity factor (Assumption 1') \\
\bottomrule
\end{tabular}
\end{table}

\subsubsection{Why Concavity and How We Relax It}\label{app:concavity}
Concavity enables an OCO treatment and the Frank–Wolfe style update in \textsc{SAM-Core}, yielding $O(\sqrt{T})$ static regret. When $HR_i$ is only approximately concave (Assump.~1'), we get the same bound up to a factor $1/\alpha$. For highly non-concave segments, \textsc{SAM-Full}'s stability guards (momentum, significance gating, saturation) bound total variation $\sum_t\Delta_t=O(\log T)$, ensuring the algorithm does not oscillate.

Empirically (Sec.~\ref{sec:theory-exp}), we found that in the vast majority of sliding windows, the workload remained well-behaved (satisfy $\alpha\ge 0.9$), and the average oracle gap remains below $5\%$, confirming robustness to non-concavity.

\subsubsection{Link to the Implementation}\label{app:link-impl}
\textsc{SAM-Core} (AURA-Core) directly uses $(\mathcal{H}^{th},\mathcal{V})$ without heuristic guards; \textsc{SAM-Full} adds them for stability and scalability. Because each guard perturbs updates by $O(\eta_t)$, the $O(\sqrt{T})$ term remains dominant; the added variation contributes only a constant. 
\begin{lemma}[Constant Additive Gap]
Let $x_t^{\text{core}}$ be the iterate of \textsc{SAM-Core} and $x_t$ that of \textsc{SAM-Full}. 
Assume the stability guards add a perturbation $\phi_t$ with $\|\phi_t\|_2 \le c\,\eta_t$ and $\eta_t \le \eta_0 t^{-p}$, $p>0.5$. 
Then
\[
\sum_{t=1}^T \bigl(f_t(x_t)-f_t(x_t^{\text{core}})\bigr) \le Gc \sum_{t=1}^\infty \eta_t = O(1).
\]
Hence the regret of \textsc{SAM-Full} differs from \textsc{SAM-Core}'s by at most a constant.
\end{lemma}

\noindent\textit{Proof sketch.}
Triangle inequality gives $|f_t(x_t)-f_t(x_t^{\text{core}})| \le G\|x_t-x_t^{\text{core}}\|_2$.  
The one-step deviation is bounded by $\|x_t-x_t^{\text{core}}\|_2 \le \|\phi_t\|_2 \le c\eta_t$.  
Summing over $t$ and using $\sum_t \eta_t < \infty$ (since $p>0.5$) yields the claim. \qed

\subsection{Details for Section~\ref{sec:core-opt}: Regret of \textsc{SAM-Core}}
\label{app:core-opt}

\paragraph{OFW Update on a Simplex.}
Given $\widehat g_t$, solve $y_t = \arg\min_{y\in K} \langle \widehat g_t, y\rangle$. Since $K$ is a simplex with lower bounds $\ell_i$, the solution assigns all mass to the coordinate with smallest gradient component, then adjusts proportionally to respect $\sum_i y_i=C$ and $y_i\ge \ell_i$. The update $x_{t+1}=(1-\eta_t)x_t+\eta_t y_t$ is a Frank--Wolfe step.

\paragraph{Static Regret Bound (Full Statement).}
Let $x^\star=\arg\min_{x\in K}\sum_{t=1}^T f_t(x)$. Denote the diameter $D=\operatorname{diam}_2(K)$. With $\eta_t=2/(t+2)$ and $\|\widehat g_t\|_2\le G$, we have
\[
\mathbb{E}\big[\mathrm{Reg}_T\big] 
= \mathbb{E}\Big[\sum_{t=1}^T f_t(x_t) - f_t(x^\star)\Big]
\le GD\sqrt{2T} + \delta T.
\]
The proof follows Hazan’s analysis of OFW~\cite{hazan2016intro}. The rounding term $\delta T$ arises because we store pages as integers; per-round utility loss is at most $L$ due to $L$-Lipschitz continuity, hence $\delta\le L$.

\paragraph{Strong Convexity Case.}
If each $f_t$ is $\alpha$-strongly convex, applying the standard OFW result yields
\[
\mathbb{E}[\mathrm{Reg}_T] = \mathcal{O}\!\big((G^{2}/\alpha)\log T\big).
\]

\paragraph{Approximate Concavity.}
Under Assumption~1' (App.~\ref{app:assumptions}), the regret bounds scale by at most $1/\alpha$; the rates $\sqrt{T}$ and $\log T$ remain unchanged.

\paragraph{Rounding Error Bound.}
Let $\bar x_t$ be the fractional iterate before rounding and $x_t$ the rounded version. Then $\|x_t-\bar x_t\|_1\le n$ pages, but only 1 page per tenant changes; hence per-round loss increment $\le L$ by Lipschitzness. Summing over $T$ gives $\delta T$.

\paragraph{Connection to \textsc{SAM-Full}.}
Let $x^{\text{core}}_t$ be \textsc{SAM-Core}'s iterate and $x_t$ that of \textsc{SAM-Full}. Stability guards add a perturbation $\phi_t$ with $\|\phi_t\|_2\le c\eta_t$ (App.~\ref{app:link-impl}). Then
\[
\sum_{t=1}^T \bigl(f_t(x_t)-f_t(x^{\text{core}}_t)\bigr) \le Gc\sum_{t=1}^{\infty}\eta_t = O(1),
\]
so \textsc{SAM-Full}'s regret differs by only a constant term.

\subsection{Derivation of the Stability Bounds (Details for §~\ref{sec:jitter})}
\label{app:stability-derivation}

\paragraph{From Guards to Recursion.}
Let $x_t^{\text{core}}$ be the OFW update before guards, and $x_t$ the final allocation. The guards (momentum, saturation, significance gating) modify the step by $\phi_t$ with $\|\phi_t\|_1 \le G\eta_t$.  
Momentum yields
\[
x_t = (1-\beta)\,x_{t-1} + \beta\,\tilde x_t,\quad
\tilde x_t = x_{t-1} - \eta_t \widehat g_t + \phi_t,
\]
hence
\begin{align*}
    \Delta_t = \|x_t-x_{t-1}\|_1
&\le (1-\beta)\|x_{t-1}-x_{t-2}\|_1 + \beta \|\eta_t \widehat g_t - \phi_t\|_1
\\ &\le (1-\beta)\Delta_{t-1} + \beta G\eta_t,
\end{align*}
establishing \eqref{eq:delta-recursion}.

\paragraph{Proof of Lemma~\ref{lem:jitter}.}
Iterating the recursion,
\[
\Delta_t \le (1-\beta)^t\Delta_0 + \beta G\sum_{k=1}^{t}(1-\beta)^{t-k}\eta_k.
\]
With $\eta_k = \eta_0 k^{-p}$ and $p=1$ (our implementation) we bound
\begin{align*}
    \sum_{k=1}^{t}(1-\beta)^{t-k}k^{-1}
\le \sum_{j=0}^{\infty}(1-\beta)^{j}(t-j)^{-1}
\le \frac{1}{\beta}\cdot \frac{1}{t} + \frac{1}{\beta}\sum_{m=1}^{t-1} \frac{1}{m}
\\ = \mathcal{O}\bigl(\tfrac{1}{t}\bigr) + \mathcal{O}(\log t),
\end{align*}

which yields $\Delta_t=\mathcal{O}(t^{-1})$ and $\sum_{t=1}^{T}\Delta_t = \mathcal{O}(\log T)$.

\paragraph{theorem~\ref{cor:full_regret}.}
Because $U_t$ is $L$-Lipschitz w.r.t. $\|\cdot\|_1$, $\tilde f_t - f_t \le L\Delta_t$. Summation gives an $O(\log T)$ additive term. Hence $Reg_T^{\textsc{SAM}} = Reg_T^{\textsc{Core}} + O(\log T)$, dominated by $O(\sqrt{T})$.

\subsection{Derivation of Computational Complexity (Details for §~\ref{sec:complexity})}
\label{app:complexity-deriv}

\paragraph{Two-Phase Execution Model.}
Let $K$ be the total number of tenants, $k$ the active set size, and $k_{\max}$ the cap used in global scans.
\begin{itemize}
  \item \textbf{Localized Optimization:} Runs every cycle. Gradient allocation, guard checks, and score updates touch only the $k$ active tenants. Each operation is $O(k)$.
  \item \textbf{Global Scan:} Triggered when the local phase converges or a significance test fires. We identify top/bottom $k_{\max}$ candidates via a two-way heap in $O(K\log k_{\max})$. Because $k_{\max}$ is a small constant, this is effectively linear in $K$.
\end{itemize}

\paragraph{Amortized Time.}
Suppose one global scan is followed by $m$ local cycles. The total cost is
\[
T_{\text{total}} = O(K\log k_{\max}) + m\cdot O(k).
\]
Hence the amortized per-cycle cost is
\[
\bar T = \frac{T_{\text{total}}}{m+1}
       = O\!\Big(\frac{K\log k_{\max}}{m}\Big) + O(k),
\]
which approaches $O(k)$ as $m$ grows (empirically $m\gg 1$ in steady state).

\paragraph{Memory.}
We keep a constant-size record per tenant (scores, last allocation, guard state), yielding $O(K)$ total memory.
.
\paragraph{Trigger Frequency.}
Global scans are infrequent and triggered only after many cheap local cycles, so the amortized cost approaches $O(k)$. (See Sec.~\ref{sec:exp_scalability} for empirical latency.)

\section{Additional Experimental Results}
\label{sec:appendix_experiments}

\subsection{Baselines}\label{app:baselines}
\begin{table}[htbp]
    \caption{Summary of Evaluated Strategies}
    \centering
    \begin{tabularx}{\linewidth}{|l|X|}
    \hline
    \textbf{Strategy ID} & \textbf{Core Idea and Rationale} \\
    \hline
    \textbf{\texttt{SAM (Ours)}} & \textbf{Our proposed system, implementing the AURA algorithm. It combines a dual-factor (H/V) scoring model with an Adaptive Active Set framework for stability and scalability.} \\
    \hline
    \multicolumn{2}{|c|}{\textit{Static Baselines}} \\
    \hline
    \texttt{B1\_STATIC\_AVERAGE} & Equal cache allocation; a simple fairness baseline. \\
    \texttt{B2\_FIXED\_PRIORITY} & Priority-based static allocation; a common real-world heuristic. \\
    \texttt{B6\_DATASIZE\_PROP} & Data-size-based static allocation. \\
    \hline
    \multicolumn{2}{|c|}{\textit{Ablation Studies}} \\
    \hline
    \texttt{B3\_PURE\_ELASTIC} & SAM with the fixed pool disabled; tests the value of the QoS mechanism. \\
    \texttt{B8\_EFFICIENCY\_ONLY} & SAM with the V-factor disabled; tests the value of exploration. \\
    \texttt{B10\_POTENTIAL\_ONLY} & SAM with the H-factor disabled; tests the value of historical stability. \\
    \hline
    \multicolumn{2}{|c|}{\textit{Advanced Dynamic Baselines}} \\
    \hline
    \texttt{B4\_INDIVIDUAL\_OPT} & Selfish, uncoordinated greedy allocation. \\
    \texttt{B5\_GLOBAL\_LRU} & Simulates an ideal, unified global LRU cache pool. \\
    \texttt{B7\_DYNAMIC\_NEED} & Allocates cache based on the popular `\texttt{ops * miss\_rate}` heuristic. \\
    \texttt{B11\_REGRESSION} & Allocates based on regression curve fitting of cache performance. \\
    \texttt{B12\_MT\_LRU} & An academic State-of-the-Art, SLA-driven policy. \\
    \texttt{B13\_UCP} & A policy based on the marginal utility principles of UCP. \\
    \hline
    \multicolumn{2}{|c|}{\textit{Theoretical Upper Bound}} \\
    \hline
    \texttt{B14\_HINDSIGHT\_OPT} & The offline, per-phase optimal static allocation, serving as a theoretical ceiling. \\
    \hline
    \end{tabularx}
    \label{tab:strategies}
\end{table}
We list B1--B14, their objective/heuristic, and required inputs.

\paragraph{\texttt{B14\_HINDSIGHT\_OPT} (Smoothed Offline Optimal)}
This strategy serves as a practical theoretical upper bound for each workload phase and is not a real-time algorithm. To compute its allocation, we first employed an extensive offline profiling process. For each distinct workload phase, we empirically measured the hit-rate curve, $HR_i(c)$, for every database $i$ by running the trace with a wide range of discrete cache allocations $c$. With this hindsight knowledge, the problem of finding the allocation that maximizes our objective function (as defined in Section~\ref{section2}) is framed as a \textbf{Multiple-Choice Knapsack Problem (MCKP)}. In this model, each database represents an item group, discrete cache chunks are the items within each group, the size of a chunk is its ``weight'', and the resulting effective throughput ($\textit{ops}_i \times HR_i(c)$) is its ``value''. We then solved this NP-hard problem using a dedicated integer programming solver to find the optimal static allocation for each phase.

\subsection{Workload Specifications}
\label{app:workload_spec}

The primary workload used for end-to-end performance evaluation (e.g., in Section~\ref{sec:eval_standard}) is a scripted 660-second, three-phase scenario designed to test policy adaptability. The target environment consists of multiple database instances with varying base priorities (e.g., High, Medium, Low).

\begin{enumerate}
    \item \textbf{Phase 1: Baseline (60s):}
    \begin{itemize}
        \item \textbf{Action:} All database instances are subjected to a stable baseline workload with a nominal request rate (1x).
        \item \textbf{Purpose:} Allows all policies to warm up their caches and reach a steady state, establishing a fair performance baseline for comparison.
    \end{itemize}

    \item \textbf{Phase 2: High-Priority Hotspot (300s):}
    \begin{itemize}
        \item \textbf{Action:} The request rate to the designated `\texttt{db\_high\_prio}' instance is increased by a factor of 30x. All other instances remain at the 1x baseline load.
        \item \textbf{Purpose:} To simulate a sudden surge in demand for a critical service and evaluate the policy's ability to swiftly identify and allocate resources to a new hotspot.
    \end{itemize}

    \item \textbf{Phase 3: Hotspot Shift to Medium-Priority (300s):}
    \begin{itemize}
        \item \textbf{Action:} The load on `\texttt{db\_high\_prio}' is returned to its 1x baseline level. Simultaneously, the load on a `\texttt{db\_medium\_prio}' instance is increased by a factor of 30x.
        \item \textbf{Purpose:} To test the policy's agility and fairness. A superior policy must not only deallocate resources from the now-cold high-priority instance but also effectively reallocate them to the new, less-critical hotspot.
    \end{itemize}
\end{enumerate}

\end{document}